\documentclass[%
 reprint,
superscriptaddress,
amsmath,amssymb,
aps,
pra,
]{revtex4-2}

\usepackage[colorlinks,linkcolor=blue,anchorcolor=blue,citecolor=blue]{hyperref}

\usepackage{float}
\usepackage{graphicx}
\usepackage{dcolumn}
\usepackage{bm}

\usepackage{braket}

\begin{document}

\preprint{APS/123-QED}

\title{Fully passive Measurement Device Independent Quantum Key Distribution}

\author{Xiang Wang}
\email{This authors contributed equally to this work.}
\author{Feng-Yu Lu}
\email{This authors contributed equally to this work.}
\author{Ze-Hao Wang}
\author{Zhen-Qiang Yin}
\email{yinzq@ustc.edu.cn}
\author{Shuang Wang}
\email{wshuang@ustc.edu.cn}

\author{Wei Chen}
\author{De-Yong He}
\author{Guang-Can Guo}
\author{Zheng-Fu Han}

\affiliation{CAS Key Laboratory of Quantum Information, University of Science and Technology of China, Hefei, Anhui 230026, China}
\affiliation{CAS Center for Excellence in Quantum Information and Quantum Physics, University of Science and Technology of China, Hefei, Anhui 230026, China}
\affiliation{Hefei National Laboratory, University of Science and Technology of China, Hefei 230088, China}



    

\date{\today}

\begin{abstract}
  Measurement-device-independent quantum key distribution (MDI-QKD) can resist all attacks on the detection devices, but there are still some security issues related to the source side. One possible solution is to use the passive protocol to eliminate the side channels introduced by active modulators at the source. Recently, a fully passive QKD protocol has been proposed that can simultaneously achieve passive encoding and passive decoy-state modulation using linear optics.
  In this work, we propose a fully passive MDI-QKD scheme that can protect the system from both side channels of source modulators and attacks on the measurement devices, which can significantly improve the implementation security of the QKD systems. 
  We provide a specific passive encoding strategy and a method for decoy-state analysis, followed by simulation results for the secure key rate in the asymptotic scenario. Our work offers a feasible way to improve the implementation security of QKD systems, and serves as a reference for achieving passive QKD schemes using realistic devices.

\end{abstract}

\maketitle


\section{\label{sec1}INTRODUCTION }

Quantum key distribution (QKD) is a communication method that uses the principles of quantum mechanics to enable the sharing of secure keys between two remote users \cite{bennett1984proceedings, ekert1991quantum}. 
Despite the fact that it is information-theoretic secure \cite{lo1999unconditional, shor2000simple, renner2008security, portmann2022security}, however, in practical applications, QKD systems may be affected by various device loopholes and channel noise \cite{scarani2009security, jain2016attacks, xu2020secure}, thus reducing their security. 
Therefore, designing a more secure, efficient, and practical QKD protocol is an essential challenge in quantum communication.

Measurement-device-independent quantum key distribution \cite{lo2012measurement, braunstein2012side} (MDI-QKD) can protect the QKD systems from any attacks on the detection side, while still having a comparable performance. The MDI-QKD has got extensive concern because it provides a balance between performance and implementation security \cite{tang2014measurement, wang2015phase, yin2016measurement, pirandola2015high, liu2019experimental,semenenko2020chip, wei2020high, comandar2016quantum, woodward2021gigahertz, fan2022robust,lu2022unbalanced}.
However, security loopholes at the source side still exist. 
For instance, Eve can launch a Trojan horse attack and extract information from the backscattered light \cite{gisinTrojanhorseAttacksQuantumkeydistribution2006, jain2014risk, jain2014trojan}. 
Moreover, in the high-speed QKD systems, the active modulators may cause correlated fluctuations in optical pulses \cite{yoshino2018quantum, roberts2018patterning, lu2021intensity}, which will violate the assumption of most security theories.
Recently, some new physical effects have been found that can attack the modulators in the transmitter and get information \cite{huang2019laser,pang2020hacking, ye2023induced, lu2023hacking}.
Previous research has proposed the passive scheme \cite{curtyNonPoissonianStatisticsPoissonian2009, curtyPassiveDecoystateQuantum2010, curtyPassiveSourcesBennettBrassard2010} to achieve encoding or decoy-state modulation \cite{hwang2003quantum,lo2005decoy,wang2005beating}.
However, this approach has a problem: the intensity and polarization of the prepared states are correlated in a passive QKD setup. Therefore, the decoy-state modulation and the encoding are hard to perform simultaneously, which limits its practical application. \cite{curty2015passive, wang2016scheme, zhang2018proof}.
It is worth noting that one new solution called fully passive QKD \cite{wangFullyPassiveQuantum2023,zapateroFullyPassiveTransmitter2023} has been proposed recently that enables both passive encoding and passive decoy-state modulation using linear optics, which makes a practical passive QKD scheme feasible.

To deal with the potential security issues at the source side of the MDI-QKD system, we aim to construct a fully passive transmitter for the MDI-QKD system that can eliminate both the side channels caused by the active modulators and attacks on the detection side, for higher implementation security. Therefore we propose a fully passive MDI-QKD scheme. 
In our scheme, Alice and Bob passively prepare arbitrary quantum states using only a linear optical structure without active modulators, and then send them to a third party for Bell state measurement. Both encoding and decoy-state modulation are performed by post-selection through the monitoring device, which can record intensity and phase information \cite{kang2023patterning,lu2023intensity}.
Furthermore, we provide a decoupling strategy to remove the correlation between intensity and polarization distribution in passive source, which allows us to use the standard decoy state analysis.
In addition, since the states prepared by the passive source are postselected in a finite region, the output signals are mixed states, so we also provide a security analysis for this mixed-state case.
With our scheme, we are finally able to achieve a QKD system with higher implementation security. Furthermore, since we remove the externally driven elements, our scheme can potentially be used in many practical scenarios.  

\begin{figure*}
  \includegraphics[width=1.0\textwidth]{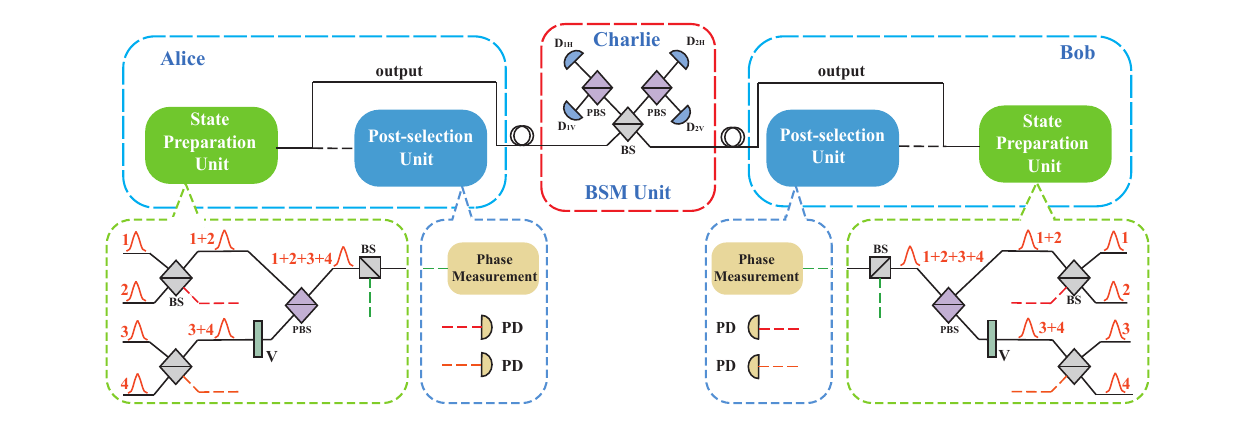}
  \caption{\label{f1}Fully Passive MDI-QKD Scheme. BS: beam splitter; PBS: polarization beam splitter; PD: photodiode. State preparation unit is used to passively and randomly generate quantum states of different intensities. Post-selection unit is employed to post-select the encoding states and decoy states. BSM unit is used to perform Bell State Measurement. In the state preparation unit, the red and orange dashed lines indicate the need to detect intensity information, and the green dashed line indicates the need to detect phase information. Take the polarization encoding structure as an example.
  For passive encoding, Alice (Bob) uses gain-switched lasers to generate four pulses with random phases, then interfere at BS, respectively. Based on the randomness of the phase, we can prepare quantum states of arbitrary intensity and arbitrary polarization. Further, we detect the intensity and phase information in each round and subsequently encode $\left\{ \rm{H},\rm{V},\rm{+},\rm{-} \right\}$ by post-selection. Alice and Bob's raw key bits would be correlated by a BSM, which could be done by an untrusted party Charlie.
  }
\end{figure*}

The paper is organized as follows. 
We propose a fully passive MDI-QKD scheme in Sec. \ref{sec2}; we provide a passive encoding strategy with a detailed carving of the post-selection regions. In addition, we give a decoupling strategy for removing the correlation between intensity and polarization in passive sources. 
In Sec. \ref{sec3}, we analyze the security issues arising from preparing mixed states in this passive scheme.
We give the numerical simulation results for the asymptotic case in Sec. \ref{sec4} and conclude the paper in the last part.

\section{\label{sec2}Fully-passive MDI-QKD }
We describe the fully passive MDI-QKD scheme in this section. Here we show our scheme with \textbf{polarization encoding as an example}, and the time-bin encoding structure can also be implemented with minor adjustments.

\subsection{Protocol}
In our scheme, we have designed a passive source structure containing a state preparation unit and a post-selection unit. The state preparation unit is used to passively generate quantum states with different intensities and different polarizations that satisfies 
\begin{equation}
  \begin{aligned}
    \ket{\psi}= c_0 \ket{0} +c_1e^{i \phi}\ket{1},\
    \ket{\psi^{\prime}}= c_0^{\prime} \ket{0}+c_1^{\prime}e^{i \phi^{\prime}}\ket{1},
  \end{aligned}
\end{equation}
where $\ket{\psi}$ and $\ket{\psi^{\prime}}$ correspond to the quantum states prepared by Alice and Bob,respectively. The coefficients $c_0= \cos (\theta/2)$ and $c_1=\sin (\theta/2)$ ($c_0^{\prime}= \cos(\theta^{\prime}/2)$ and $c_1^{\prime}=\sin (\theta^{\prime}/2)$)  determine the specific form of the encoding states on Bloch sphere, where $\theta$ and $\theta^{\prime}$ are the polar angles, $\phi$ and $\phi^{\prime}$ are the azimuth angles.
Subsequently, the post-selection unit locally detects the intensity and phase information for post-selection to achieve passive encoding and decoy-state modulation. Figure \ref{f1} gives the specific structure of the fully passive MDI-QKD scheme and indicates how the passive source structure works. 

Alice (Bob) randomly prepares quantum states of different intensities and polarizations in each turn and then post-selects the encoding and decoy states. These are then sent to Charlie for the Bell state measurement (BSM). Charlie publicly announces whether or not a successful measurement event has been obtained.
The following outline describes the protocol procedure:

\emph{1.Preparation.} In each turn, Alice (Bob) randomly prepares a quantum state $\ket{\psi}$ ($\ket{\psi^{\prime}}$) and sends it to the measurement unit Charlie. They will record the intensity and phase information of the prepared state through the post-selection unit, which determines the basis Z (or X) and bit 0 (or 1). Encoding and decoy-state modulation will be performed simultaneously through post-selection.

\emph{2.Measurement.} Charlie projects the received photon pairs to the bell state $\left|\psi^{+}\right\rangle=\frac{1}{\sqrt{2}}\left(|0\rangle|1\rangle+|1\rangle|0\rangle\right)$ and $\left|\psi^{-}\right\rangle=\frac{1}{\sqrt{2}}\left(|0\rangle|1\rangle-|1\rangle|0\rangle\right)$.
$\ket{\psi^+}$ corresponds to coincident detections of $D_{\rm{1H}} \& D_{\rm{1V}}$ or $D_{\rm{2H}} \& D_{\rm{2V}}$, and 
$\ket{\psi^-}$ corresponds to coincident detections of $D_{\rm{1H}} \& D_{\rm{2V}}$ or $D_{\rm{1V}} \& D_{\rm{2H}}$.
Alice and Bob generate the raw key based on the published results.

\emph{3.Sifting.} After the above steps have been repeated enough times, Alice (Bob) publicly announces the selection of basis and decoy intensity for each round. When both users select the same basis and Charlie announces a valid response, they can extract the raw key.

\emph{4.Parameter estimation.} By choosing an appropriate post-selection region and reshaping the intensity probability distribution, the intensity can be decoupled from the polarization, though sacrificing some data. 
This enables a linear program to estimate the single photon yield and single photon bit error rate.

\emph{5.Error correction and privacy amplification.} Alice and Bob perform error correction and privacy amplification to obtain the secure key based on the final parameter estimation results.

In the fully passive MDI-QKD scheme, there is an intensity probability distribution of the output state,
and we encode the states by dividing the intensity regions. Thus the observable quantity $Q$ will turn out to be an expectation value, determined by the selected post-selection regions.
Then, after applying the fully passive source, the secure key rate should be:
\begin{eqnarray}
  {\cal R} \geq &&P_{S_{\chi}^{\rm{Z}}} P_{S_{\chi^{\prime}}^{\rm{Z}}} \left\{ \langle P_{11} \rangle _{S_{\chi}^{\rm{Z}}S_{\chi^{\prime}}^{\rm{Z}}} Y_{11}^{\rm{Z},L} [1-H(e_{11}^{\rm{X},U})]\right. \nonumber\\
  &&\left. -f_e \langle Q \rangle _{S_{\chi}^{\rm{Z}}S_{\chi^{\prime}}^{\rm{Z}}} H({\langle T \rangle _{S_{\chi}^{\rm{Z}}S_{\chi^{\prime}}^{\rm{Z}}}}/{\langle Q \rangle _{S_{\chi}^{\rm{Z}}S_{\chi^{\prime}}^{\rm{Z}}}}) \right\}~. 
  \label{e2}
\end{eqnarray}  
where $Y_{11}^{\rm{Z},L}$ and $e_{11}^{\rm{X},U}$ are the lower bound of single photon yield in the Z basis and the upper bound of single photon error rate in the X basis.
Here we use the Z basis for key generation and the X basis for parameter estimation only.
$S_{\chi}^{\rm{Z}}$ ($S_{\chi^{\prime}}^{\rm{Z}}$) denotes the maximum post-selection region used by Alice (Bob) to select Z basis (Sec. \ref{B2} gives a more specific definition of the post-selection regions).
$P_{S_{\chi}^{\rm{Z}}}$ and $P_{S_{\chi^{\prime}}^{\rm{Z}}}$ are the sifting probabilities for choosing the key generation region respectively.
$\langle P_{11} \rangle _{S_{\chi}^{\rm{Z}}S_{\chi^{\prime}}^{\rm{Z}}}$ is the probability that Alice and Bob send a single photon simultaneously in the Z basis.
$H(x)=-x\log_2(x)-(1-x)\log_2(1-x)$ is the binary entropy function, and $f_e$ is the error correction coefficient. 
$\langle Q \rangle _{S_{\chi}^{\rm{Z}}S_{\chi^{\prime}}^{\rm{Z}}}$ and $\langle T \rangle _{S_{\chi}^{\rm{Z}}S_{\chi^{\prime}}^{\rm{Z}}}$ denote, respectively, the gain and the error gain in Z basis.

\subsection{\label{B2}Passive encoding strategy}
Alice (Bob) used a passive source to prepare quantum states with different intensities and polarizations, and we need to give a specific post-selection strategy to select valid data points from the intensity region and phase region for base selection and encoding. Before that, we first analyze how the passive source works.

In Fig. \ref{f1}, Alice uses a phase-randomized laser to generate four strong optical pulses with a uniform phase distribution between $[0,2\pi)$:
\begin{eqnarray}
 \ket{\sqrt{\mu_{\rm{in}}} e^{i \phi_1}}_1 \ket{\sqrt{\mu_{\rm{in}}} e^{i \phi_2}}_2 
 \ket{\sqrt{\mu_{\rm{in}}} e^{i \phi_3}}_3 \ket{\sqrt{\mu_{\rm{in}}} e^{i \phi_4}}_4 ,
\end{eqnarray}
where $\mu_{\rm{in}}$ is the output intensity of the laser and $\phi_1$ to $\phi_4$ are the phases of the four pulses.
Alice sends the four pulses to the passive encoding structure, where $\ket{\sqrt{\mu_{\rm{in}}} e^{i \phi_1}}$ and $\ket{\sqrt{\mu_{\rm{in}}} e^{i \phi_2}}$ interfere with each other, 
as well as $\ket{\sqrt{\mu_{\rm{in}}} e^{i \phi_3}}$ and $\ket{\sqrt{\mu_{\rm{in}}} e^{i \phi_4}}$, 
to produce two intermediate coherent states $\ket{\sqrt{\mu_{H}} e^{i \phi_{H}}}_{H}$ and $\ket{\sqrt{\mu_{V}} e^{i\phi_{V}}}_{V}$. 
They are then combined at the PBS and the output state can be write as $\ket{\sqrt{\mu} e^{i \phi_{G}}}_{\theta, \phi}$. The intensity $\mu$, polar angle $\theta$ and azimuthal angle $\phi$ uniquely determine a single photon state on the bloch sphere, with a global random phase $\phi_{G}$, 
which can be abtained as
\begin{equation}
  \begin{aligned}
    &\mu = \mu_H +\mu_V \\
    &\theta = 2\arccos\sqrt{\mu_{H}/(\mu_{H}+\mu_{V})} \\
    &\phi = \phi_{V}-\phi_{H} \\
    &\phi_G = \phi_H,
  \end{aligned}
\end{equation}
where $\mu_{H}=\mu_{in}[1+\cos(\phi_2-\phi_1)]$ and $\mu_{V}=\mu_{in}[1+\cos(\phi_4-\phi_3)]$, 
$\phi_H=\left(\phi_1+\phi_2\right) / 2$ and $\phi_V=\left(\phi_3+\phi_4\right) / 2$.
Bob follows the same procedure to prepare the state, obtaining $\ket{\sqrt{\mu^{\prime}} e^{i \phi_{G}^{\prime}}}_{\theta^{\prime}, \phi^{\prime}}$.
Similarly, the quantities $\mu^{\prime}$, $\theta^{\prime}$, $\phi^{\prime}$ and $\phi_{G}^{\prime}$ can be abtained from Eq.(\ref{e2}).
We set the maximum intensity of the states prepared in each turn to be $\mu_{\rm{max}}=2\mu_{\rm{in}}\eta_F$, where $\eta_F$ is the total attenuation of the passive source structure. 
Note that during each round of preparation, Alice (Bob) uses the local photodiodes and phase measurement structure in the post-selection unit to get the intensity and phase information, for the next encoding step.

\begin{figure}[htb]
  \includegraphics[width=1.0\linewidth]{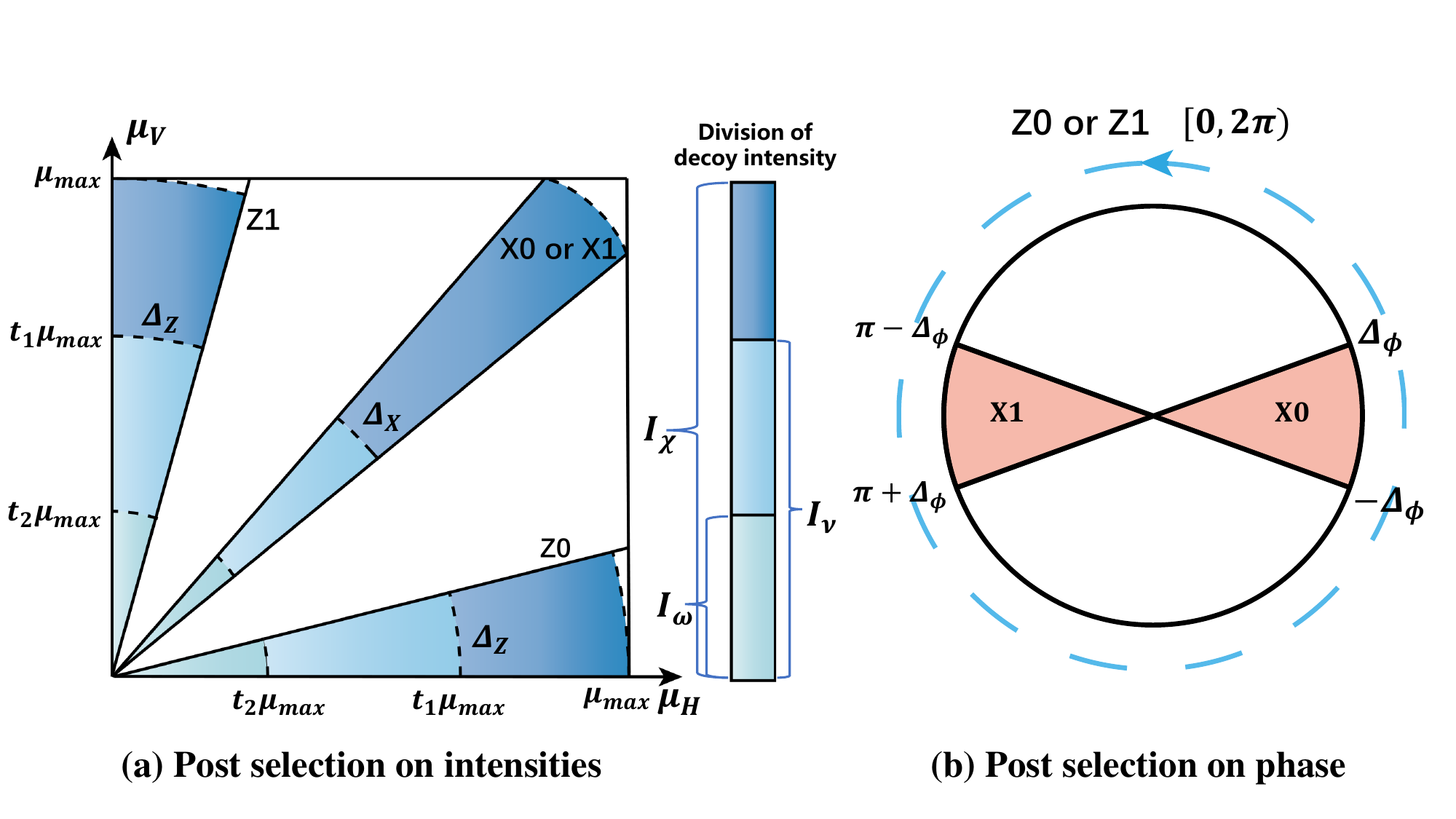}
  \caption{\label{fig2} 
  Post-selection regions for encoding states and decoy states. Alice (Bob) pre-decides the maximum intensity $\mu_{max}$ to generate an intensity space. The regions Z0 and Z1 (X0 and X1) define the choice of Z (X) basis.
  (a) Post selection on intensity $\mu$. The horizontal (vertical) axis denotes the intensity $\mu_H$ ($\mu_V$), and the total intensity of any point in the figure is $\mu=\mu_H+\mu_V$.
  $\Delta_Z$ and $\Delta_X$ show the range of the encoding regions for Z basis and X basis, respectively, which also determine the threshold of the polar angles on the Bloch sphere. 
  The $t_1$ and $t_2$ determine the intervals of the decoy states.
  (b) Post selection on phase $\phi$. $\Delta_{\phi}$ determines the threshold of the azimuth angles on the Bloch sphere. The azimuth angles locate in pink parts represent the encoding of X0 and X1 respectively, and the phase interval is chosen as $[0,2\pi)$ for Z basis.
  }
\end{figure}

Based on the measured intensity and phase information, Alice (Bob) can define the post-selection regions to determine the basis, key bit and decoy intensity. 
As illustrated in the Fig. \ref{fig2}, Alice pre-decides the key parameters $\Delta_Z$, $\Delta_X$ and $\Delta_{\phi}$ to divide the intensity and phase regions. Define $\theta_{\mu} = \arctan (\mu_V/\mu_H)$, $r_{\mu}=\sqrt{\mu_H^2+\mu_V^2}$, when $0 \leq \theta_{\mu} \leq \Delta_Z$ ($\pi/2-\Delta_Z \leq \theta_{\mu} \leq \pi/2$), Alice post-selects the states in Z basis with bit value 0 (1). 
When $\pi/4-\Delta_X \leq \theta_{\mu} \leq \pi/4+\Delta_X$, Alice post-selects the states in X basis, and the phase satisfies $-\Delta_{\phi} \leq \phi \leq \Delta_{\phi}$ ($\pi-\Delta_{\phi} \leq \phi \leq \pi+\Delta_{\phi}$) determining the bit value 0 (1). Signals not in the above areas will be discarded.

After determining the bases and bit values, Alice (Bob) can further delineate the decoy state regions, see Fig. 2.(a). Alice pre-decides the decoy parameters $t_1$ and $t_2$, then, in terms of the maximum intensity $\mu_{\rm{max}}$, the decoy intervals are delimited by $I_{\chi}$ (i.e. $r_{\mu} \leq \mu_{\rm{max}}$), $I_{\nu}$ (i.e. $r_{\mu} \leq t_1\mu_{\rm{max}}$) and $I_{\omega}$ (i.e. $r_{\mu} \leq t_2\mu_{\rm{max}}$).
Specific to different bases only need to add restrictions on $\theta_\mu$.
Note that we choose concentric overlapping sector regions as decoy state settings, 
which has no effect on the linear programming \cite{zhou2016making} results, but has the advantage of allowing more data in each decoy state regions.

In summary, we have defined the post-selection regions $S_{i,k}^{\Omega}$ (capable of being divided by $\mu_H$, $\mu_V$ and $\phi$), where $i \in \left\{\chi, \nu, \omega \right\}$ represents the decoy intensity, $k \in \left\{0,1\right\}$ represents the key value and $\Omega \in \left\{\rm{Z},\rm{X}\right\}$ represents the basis.
Bob can define similar post-selection regions $S_{j,k}^{\Omega}$ (capable of being divided by $\mu_H^{\prime}$, $\mu_V^{\prime}$ and $\phi^{\prime}$), $j \in \left\{\chi^{\prime}, \nu^{\prime}, \omega^{\prime} \right\}$. In addition, for later convenience we define $S_{i}^{\Omega} = S_{i,0}^{\Omega} \cup S_{i,1}^{\Omega}$ and $S_{j}^{\Omega} = S_{j,0}^{\Omega} \cup S_{j,1}^{\Omega}$.
Since the encoding is associated with continuous post-selection regions, the intensity of the prepared states are correlated with the polarization in this passive source. Therefore, in order to perform a standard decoy state analysis, users must implement a decoupling strategy to remove the correlation between intensity and polarization distribution.

\subsection{\label{sec4-2}Decoupling strategy}
In the passive MDI-QKD system case, the decoy states are not discrete intensity points, 
but continuous intensity regions $S_i^{\Omega}$ and $S_j^{\Omega}$ that we get from post-selecting on the intensity probability distribution. 
For simplicity, we omit the basis-vector symbol $\Omega$ in the following presentation, and our discussion applies for both Z basis and X basis.
In a fully passive MDI-QKD implementation, by performing the measurements for different decoy intensity settings, we can obtain 
\begin{equation}
  \begin{aligned}
    \langle Q \rangle _{S_i S_j}=\sum^{\infty}_{n,m=0}\langle P_{nm}Y_{nm} \rangle _{S_i S_j},\\
    \langle T \rangle _{S_i S_j} =\sum^{\infty}_{n,m=0}\langle P_{nm}E_{nm}\rangle _{S_i S_j},
  \end{aligned}
  \label{e14}
\end{equation}
where $P_{nm}$, $Y_{nm}$ and $E_{nm}$ are respectively the joint Poissonian distribution, the yield, and the error yield, while Alice sending an $n$-photon pulse and Bob sending an $m$-photon pulse. 
$P_{nm}$ is a function of $(\mu_H,\mu_V,\mu_H^{\prime},\mu_V^{\prime})$, while $Y_{nm}$ and $E_{nm}$ are functions of $(\mu_H,\mu_V,\phi,\mu_H^{\prime},\mu_V^{\prime},\phi^{\prime})$.
$E_{nm}=e_{nm}Y_{nm}$, where $e_{nm}$ denotes the quantum bit error rate (QBER). 
Here, the key idea of the finite decoy-state protocol is to estimate $Y_{11}^{\rm{Z,L}}$ and $e_{11}^{\rm{X,U}}$ from the set of linear equations given by Eq.(\ref{e14}). 
For an arbitrary function $f(\mu_H,\mu_V,\phi,\mu_H^{\prime},\mu_V^{\prime},\phi^{\prime})$, its expected value on the post-selection regions $S_i$ and $S_j$ can be written as
\begin{equation}
  \begin{aligned}
  & \left\langle f(\mu_H, \mu_V, \phi,\mu_H^{\prime},\mu_V^{\prime},\phi^{\prime}) \right\rangle _{S_i S_j} \\
  = & \frac{1}{P_{S_i S_j}} \iiint \iiint_{S_i S_j} p\left(\mu_H, \mu_V,\phi\right) p\left(\mu_H^{\prime},\mu_V^{\prime},\phi^{\prime}\right) \\
  & \times f(\mu_H,\mu_V,\phi,\mu_H^{\prime},\mu_V^{\prime},\phi^{\prime}) d \mu_H d \mu_V d \phi d \mu_H^{\prime} d \mu_V^{\prime} d \phi^{\prime},
  \end{aligned}
\end{equation}
where $p\left(\mu_H, \mu_V,\phi\right)$ ($p\left(\mu_H^{\prime},\mu_V^{\prime},\phi^{\prime}\right)$) is the natural probability distribution from the passive source, $P_{S_iS_j}$ is the probability of bases selection and decoy settings
\begin{equation}
  \begin{aligned}
    P_{S_iS_j}=&\iiint \iiint_{S_iS_j} p\left(\mu_H, \mu_V,\phi\right) \times \\
    &p\left(\mu_H^{\prime},\mu_V^{\prime},\phi^{\prime}\right) 
    d \mu_H d \mu_V d \phi d \mu_H^{\prime} d \mu_V^{\prime} d \phi^{\prime}.
  \end{aligned}
\end{equation}

Note that $P_{nm}$ and $Y_{nm}$ are coupled in Eq.(\ref{e14}), as both $P_{nm}$ and $Y_{nm}$ takes a range of values in $S_i$ and $S_j$.
This means that it is not possible at this moment to find a set of continuous variables independent of the choice of the decoy state to perform linear programming.
To solve this problem, it is necessary to decople $Y_{nm}$ and $P_{nm}$, meaning that the photon number distribution $P_{nm}$ is related to the setting of the decoy state, but independent of $Y_{nm}$.

The natural probability distribution from the passive source, 
$p\left(\mu_H, \mu_V, \phi\right)=p_{\mu}\left(\mu_H, \mu_V\right) p_{\phi}\left(\phi\right)$, 
representing the probability of a given combination $\left\{ \mu_H,\mu_V,\phi \right\}$.
The specific form can be written as:
\begin{eqnarray}
  &&p_\mu\left(\mu_H, \mu_V\right)  =\frac{1}{\pi^2 \sqrt{\mu_H\left(\mu_{\max }-\mu_H\right) \mu_V\left(\mu_{\max }-\mu_V\right)}}, \nonumber\\
  &&p_\phi\left(\phi\right)  =\frac{1}{2 \pi}. 
  \label{e8}
\end{eqnarray}
Here we implement the decoupling strategy, i.e., to shape the intensity and phase probability distribution $p\left(\mu_H, \mu_V, \phi \right)$ of the source. In other words, the users keep or discard the signals with probability $q_\mu(\mu_H,\mu_V)$. This allows to obtained an arbitrary intensity probability distribution $p_\mu^{\sim}=p_{\mu}q_\mu$ they wanted.

In our scheme, we prove that if we use the post-selection regions in Fig. \ref{fig2} and shape the intensity distribution into $p_{\mu}^{\sim} \propto e^{(\mu_H+\mu_V)}$, the multiphoton term in Eq.(\ref{e14}) can be decoupled into $\langle P_{nm} \rangle_ {S_i S_j}$ and $Y_{nm}^{\prime}$, where $Y_{nm}^{\prime}$ is independent of the region selection $\left\{S_i, S_j\right\}$. See Appendix \ref{Appendix A} for more details.
A new linear programming can therefore be constructed as follows:
\begin{equation}
  \begin{aligned}
    \langle Q \rangle_ {S_iS_j}=\sum_{n,m} \langle P_{nm} \rangle _{S_iS_j} \times Y_{nm}^{\prime}, \\
    \langle T \rangle_ {S_iS_j}=\sum_{n,m} \langle P_{nm} \rangle _{S_iS_j} \times E_{nm}^{\prime}.
    \label{e11}
  \end{aligned}
\end{equation}
Here we have a new yield $Y_{nm}^{\prime}$ and an error yield $E_{nm}^{\prime}$.
And we also have proven that $Y_{nm}^{\prime}$ is the same for all decoy state settings.
So from the Eq.(\ref{e11}) we can use linear programming to obtain a lower bound of $Y_{nm}^{\prime}$ and an upper bound of $e_{nm}^{\prime}$.

Indeed, the goal of the decoy state analysis is to obtain $Y_{11}^{\rm{Z},L}$ and $e_{11}^{\rm{X},U}$ with perfectly prepared single photon, hence we will show that the yield and error yield of single photon states with perfect polarizations also have lower and upper bounds given by $Y_{nm}^{\prime}$ and $e_{nm}^{\prime}$. 
We can use the security analysis conclusion in Sec \ref{sec3}. In fact the output states prepared by the passive source are mixed states, i.e. the mixture of a pair of states respectively having a misalignment of $-\theta$ and $\theta$ from the polar angles in perfect encoding case. And this imperfect state preparation is equivalent to preparing the pure state signal normally and then adding random noise with a certain probability.
On the one hand, the single photon yield $Y_{11}$ is independent of the polarizations, thus the single photon yield in the perfect preparation case satisfies
\begin{equation}
  \begin{aligned}
    Y_{11}^{\text {perfect }}=Y_{11}^{\prime} \geq Y_{11}^{\prime, L},
  \end{aligned}
\end{equation}
which means that $Y_{11}^{\prime, L}$ is also a lower bound for $Y_{11}^{\text {perfect }}$.
On the other hand, for error rate, the average QBER for the mixed single photons cannot be smaller than the QBER of the perfectly prepared states. For every $\theta_1$ and $\theta_2$, we will have
\begin{equation}
  \begin{aligned}
    E_{11}^{\text {perfect }} \leq E_{11}^{\prime} = e_{11}^{\prime}Y_{11}^{\prime}(\theta_1,\theta_2).
  \end{aligned}
\end{equation}
Therefore, we obtain
\begin{equation}
  \begin{aligned}
    E_{11}^{\text {perfect }} \leq E_{11}^{\prime} \leq E_{11}^{\prime, U}
  \end{aligned}
\end{equation}
which means that the upper bound for the $e_{11}^{\prime}$ is also an upper bound for the $e_{11}^{\text {perfect }}$. 
And then by solving constrained optimization problems (see Appendix \ref{Appendix C}), we can obtain the estimations on $Y_{11}^{\rm{Z},L}$ and $e_{11}^{\rm{X},U}$ from Eq.(\ref{e2}).

\section{\label{sec3}security analysis }
In the fully passive MDI-QKD scheme, the main difference from the active scheme comes from the quantum state preparation process, where the quantum states prepared by both users are no longer pure states but postselected mixed states. In this section, we show that this state preparation process is equivalent to Alice's (Bob's) normal preparation of pure states, but with noise randomly added to the original key bits. In other words, Alice (Bob) randomly flips the original key bits with a certain probability. 

Our protocol will use the Z-basis signals for key generation. A similar discussion can be used for the case of using X-basis or Y-basis for key generation. Since we are passively encoding by post-selection, 
the actual states are mixed states $\rho_{H}$ and $\rho_{V}$.
We observe the fact that based on the post-selection regions we have defined, the polarization fluctuation in the source is symmetric, i.e. the distribution of the polar angles on Bloch sphere is centered at the angles of perfectly prepared case ($0$ for the $\ket{H}$ and $\pi$ for the $\ket{V}$). In the prepare-and-measure picture, when the users want to prepare pure state $\ket{H}$, due to polarization fluctuation will actually prepare 
\begin{equation}
  \begin{aligned}
    \ket{H(\theta)} = \cos (\theta/2) \ket{H} + \sin (\theta/2) \ket{V}.  \\
  \end{aligned}
\end{equation}
The polar angle $\theta$ can satisfy an arbitrary distribution $p_\theta (\theta)$, but need to obey
\begin{equation}
  \begin{aligned}
    p_\theta (-\theta) = p_\theta (\theta).
  \end{aligned}
\end{equation}
Now, due to the symmetry of the distribution, we can always find a pair of misaligned states ($\ket{H(\theta)}$ and $\ket{H(-\theta)}$ ) with equal probability. Therefore the average state between the two states would be
\begin{equation}
  \begin{aligned}
    \rho_H = \cos^2\frac{\theta}{2}\ket{H}\bra{H} + \sin^2\frac{\theta}{2}\ket{V}\bra{V},
  \end{aligned}
\end{equation}
which means the users perfectly prepare the pure state while randomly flipping the Z basis bits with a probability of $\sin^2\frac{\theta}{2}$.

In the actual passive source, Alice (Bob) randomly prepare 
\begin{equation}
  \begin{aligned}
    \ket{\psi(\theta,\phi)} = \cos \frac{\theta}{2} \ket{H} + e^{i\phi}\sin \frac{\theta}{2} \ket{V},
  \end{aligned}
\end{equation}
where $\theta$ and $\phi$ determined by post-selection regions, $p\left(\theta, \phi\right)$ is the probability density function. For Z-basis signals, the distribution of the states are three-dimensional solid angles in the “polar” regions of the Bloch sphere. Therefore, Alice actually prepares mixed states
\begin{equation}
  \begin{aligned}
  \rho_H  =\int_0^{\Delta_z} &\int_0^{2 \pi} p\left(\theta, \phi\right)  \times  \\
   &(\cos^2\frac{\theta}{2}\ket{H}\bra{H} + \sin^2\frac{\theta}{2}\ket{V}\bra{V}) 
   d \theta d \phi , \\
  \rho_V  =\int_{\pi-\Delta_Z}^\pi &\int_0^{2 \pi} p\left(\theta, \phi\right)  \times   \\
  &(\sin^2\frac{\theta}{2}\ket{H}\bra{H} + \cos^2\frac{\theta}{2}\ket{V}\bra{V}) d \theta d \phi .
  \end{aligned}
\end{equation}

It is worth noting that as long as the distributions of the polarization fluctuations in the mixture are symmetric, in other words, the probability distribution is symmetric, 
i.e. $p\left(\theta, \phi\right) = p\left(\theta, \phi+\pi \right)$, 
it is always possible to obtain
\begin{equation}
  \begin{aligned}
    \rho_H & = (1-\xi)\ket{H}\bra{H} + \xi\ket{V}\bra{V}, \\
    \rho_V & = (1-\xi)\ket{V}\bra{V} + \xi\ket{H}\bra{H},
  \end{aligned}
\end{equation}
where coefficient $\xi$ characterise the imperfection properties of the preparation, that is, the probability of random bit flips.

Therefore, this imperfect state preparation of the Z basis signals is equivalent to preparing the pure state signal normally and then adding random noise with a certain probability that this noise will not allow Eve to steal any more information. Furthermore, the upper and lower bounds on the yield and QBER of a perfectly encoded single photon can be obtained by performing decoy state analysis.

\section{\label{sec4}SIMULATION }
Based on the post-selection regions in Sec \ref{sec2}, the expected value of any observable $Q$ is an integration over the post-selection regions $S_i$ and $S_j$
\begin{equation}
  \begin{aligned}
      \langle Q\rangle _{S_iS_j}=
      (1 / P_{S_i S_j}) \int...\int_{S_i,S_j}  
      p(\boldsymbol{s}) \times p(\boldsymbol{s^{\prime}}) \\
       \times Q(\boldsymbol{s},\boldsymbol{s^{\prime}})
      \ d(\boldsymbol{s},\boldsymbol{s^{\prime}}) ,  \\
      P_{S_i S_j}  =\int...\int_{S_i,S_j} p(\boldsymbol{s}) \times p(\boldsymbol{s^{\prime}})
      \ d(\boldsymbol{s},\boldsymbol{s^{\prime}}) .
      \label{e18}
  \end{aligned}
\end{equation}
where $\boldsymbol{s}= (\mu_{H}, \mu_{V},\phi)$ ($\boldsymbol{s^{\prime}} = (\mu_{H}^{\prime}, \mu_{V}^{\prime},\phi^{\prime})$) denotes the variables that determine the post-selection regions of Alice (Bob).
$P_{S_i S_j}$ is the normalization factor, representing the probability of bases selection and decoy settings. $p(\boldsymbol{s})$ ($p(\boldsymbol{s^{\prime}})$) is the probability distribution of the output state of Alice (Bob).
$Q(\boldsymbol{s},\boldsymbol{s^{\prime}})$ defines the gain when Alice and Bob send a specific intensity point, respectively, for every given set of $\left\{ \boldsymbol{s},\boldsymbol{s^{\prime}} \right\}$. Appendix \ref{Appendix B} provides detailed definitions for $Q(\boldsymbol{s},\boldsymbol{s^{\prime}})$.

\begin{figure}[htb]
  \includegraphics[width=1.0\linewidth]{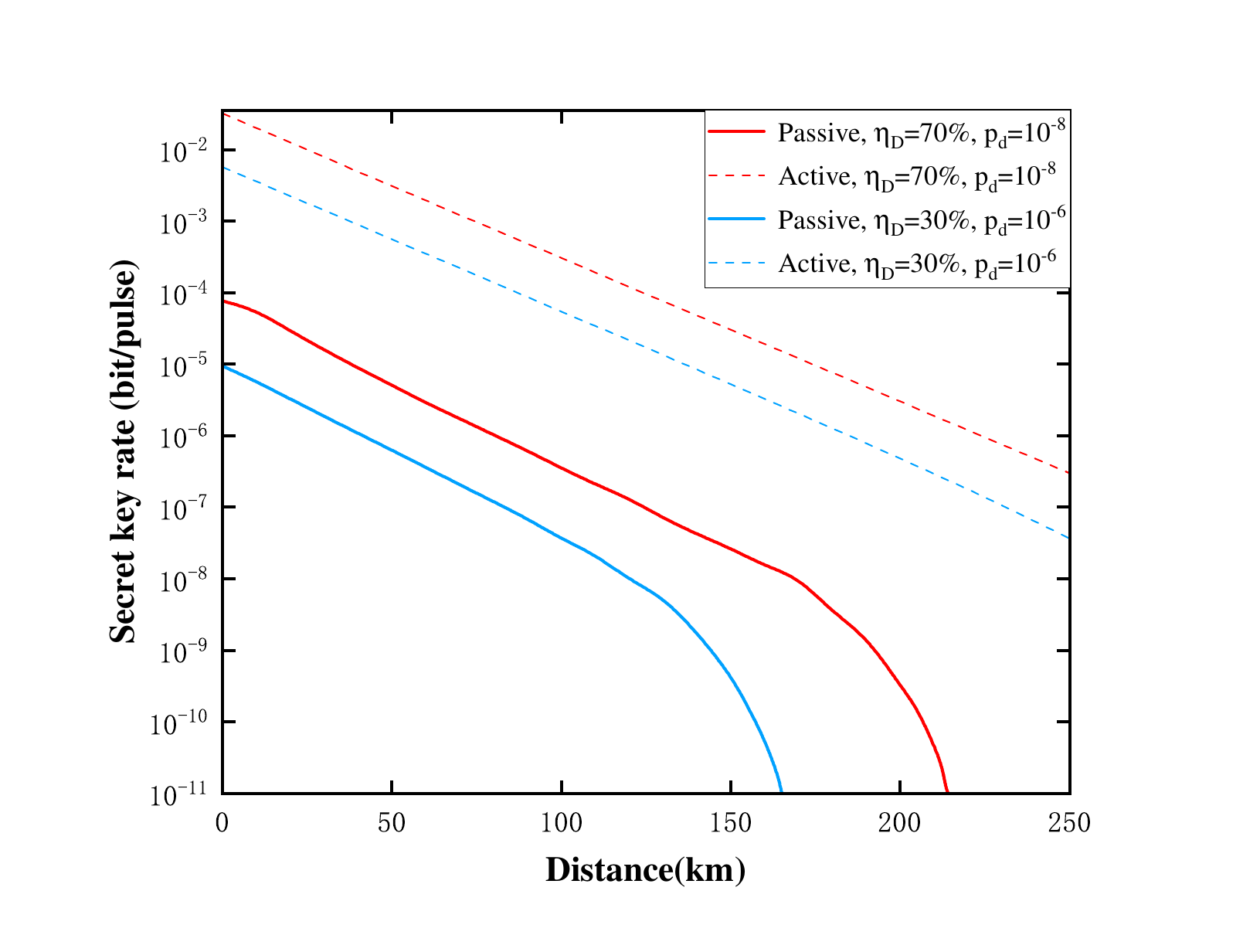}
  \caption{Simulation of the secret key rate for fully passive decoy-state MDI-QKD versus active three-intensity decoy-state MDI-QKD \cite{yu2013three}. We use the encoding strategy in Fig. \ref{fig2} with optimization for maximum intensities and the size of post-selection regions. The dashed lines represent the SKR of the active three-intensity MDI protocol, while the solid lines represent the SKR of our fully passive MDI protocol.
  In addition, we give the performance of the protocol in different scenarios. The red lines indicate the use of SNSPDs, with $\eta_D=70\%$ and $p_d=10^{-8}$, and the blue lines represent the use of SPADs, with $\eta_D=30\%$ and $p_d=10^{-6}$.
  \label{fig3}}
\end{figure}

In what follows, we give a simulation of the fully passive MDI-QKD in the asymptotic case. 
Based on the channel model we have given in Appendix \ref{Appendix B}, the overall efficiencies of the system $\eta_A=\eta_D10^{-\alpha L_A/10}$, $\eta_B=\eta_D10^{-\alpha L_B/10}$, $\alpha$ denotes the attenuation coefficient of the channel, $\eta_D$ is the detection efficiency on both sides of the users, $L_A$ ($L_B$) stands for the distance between Alice (Bob) and Charlie.
For illustration purposes, we use a symmetric MDI-QKD system, and we set the system parameters as follows: dark count rate $p_d = 1\times 10^{-8}$ and detection efficiency $\eta_D=70\%$ for superconducting nanowire single photon detectors (SNSPDs), dark count rate $p_d = 1\times 10^{-6}$ and detection efficiency $\eta_D=30\%$ for single photon avalanche photodiodes (SPADs), fiber loss $\alpha=0.2 \rm{dB/km}$, and error correction factor $f_e=1.16$.
In order to maximize the secret key rate for each value of $L$, we numerically optimize the maximum intensity $\mu_{\rm{max}}$ and the angular widths $\Delta_Z$, $\Delta_X$ and $\Delta_{\phi}$ of the post-selection regions.
Also, we numerically optimize the value of the decoy state parameter $t_{1}$ and $t_{2}$. 
Finally, we set the threshold photon number for the decoy-state linear programs to $N_{\text {cut }}=6$ and $M_{\text {cut }}=6$. Here we only consider the scenario of infinite data size. 
The results are shown in Fig. \ref{fig3}, where we further include the secret key rate reached in the active three-intensity decoy-state MDI-QKD for comparison purposes.

Figure \ref{fig3} shows that the implementation security improvement provided by the passive scheme comes at the cost of reducing the secret key rate. The passive scheme has a lower asymptotic key rate than the active scheme, about two orders of magnitude less. 
There are two main reasons for the decrease in the key rate.
On the one hand, in our scheme where Z-basis data is chosen for encoding, Alice and Bob need to perform post-selection sifting simultaneously, which will discard data outside the encoding region. On the other hand, since the post-selection regions are of finite size, the source's inherent noise increases the bit error rate.

\section{conclusion}
Fully passive schemes have been proposed to close the side channels of the active modulators at the source in the QKD systems.
In this paper, we presented a fully passive MDI-QKD scheme that could remove the security issues introduced by modulators in MDI-QKD systems. Our solution can be applied to polarization and phase encoding systems with simple adjustments.
Combined with our paper's passive encoding strategy and decoupling strategy, we can perform the standard decoy state analysis. 
Finally, we obtain the simulation results of the secure key rate for this protocol.
However, our scheme requires the users to perform an auxiliary post-selection to decouple the intensity and the polarization of the output states, which leads to more data loss. Thus, future work is to find a feasible strategy to minimize data filtering.
In conclusion, our scheme improves the implementation security of QKD systems and provides a reference for the realization of passive QKD schemes with realistic devices.

\begin{acknowledgments}
  This work has been supported by the National Natural Science Foundation of China ( No. 62271463, 62171424, and 62105318), the China Postdoctoral Science Foundation (2022M723064,2021M693098); and the Anhui Initiative in Quantum Information Technologies.
\end{acknowledgments}

\appendix
\section{\label{Appendix A}Decople of polarization and intensity}
For simplicity, we discuss this problem in polar coordinates, i.e. transforming $(\mu_H,\mu_V)$ to $(r_1,\theta_1)$ and $(\mu_H^{\prime},\mu_V^{\prime})$ to $(r_2,\theta_2)$, where 
\\
\begin{equation}
  \begin{aligned}
    r_1=\sqrt{\mu_H^2+\mu_V^2}, \ \theta_1=\arctan(\mu_V/\mu_H), \\
    r_2=\sqrt{\mu_H^{\prime 2}+\mu_V^{\prime 2}}, \ \theta_2=\arctan(\mu_V^{\prime}/\mu_H^{\prime}).
  \end{aligned}
\end{equation}
Based on the post-selection regions defined in Fig. \ref{fig2}, the decoy-state settings has the same angular integration region $(\theta_{min},\theta_{max})$, differing only in the different radial integration regions. Here we choose  concentric overlapping sector regions $[0,r_{max,S_i(j)}]$ as decoy-state settings, which can allowing more data in each $\left\{ S_i,S_j\right\}$ \cite{zhou2016making}.
Noting that the phase distribution $p(\phi)$ is independent of the intensity distribution $p_{\mu}(\mu_H,\mu_V)$, we can always integrate the phase $\phi$ separately, so we ignore the phase in the following analyses.

The specific form in Eq.(\ref{e14}) can be writte as:
\begin{widetext}
\begin{equation}
  \begin{aligned}
    \langle P_{nm} Y_{nm} \rangle_{S_iS_j}=\frac{1}{P_{S_i S_j}} \iiiint_{S_iS_j} p_{\mu}(r_1, \theta_1) p_{\mu^{\prime}}(r_2, \theta_2) P_{nm}(r_{1},r_2,\theta_1,\theta_{2}) Y_{nm}(\theta_{1},\theta_2) \ r_1 r_2 dr_1 dr_2 d \theta_1 d \theta_2,
  \end{aligned}
  \label{eA1}
\end{equation}
\end{widetext}
where $p_{\mu}(r_1, \theta_1)$ and $p_{\mu^{\prime}}(r_2, \theta_2)$ are the inherent intensity distribution from users' source, $P_{nm}(r_{1}, r_{2}, \theta_{1}, \theta_{2})$ is the joint Poissonian distribution while Alice sends n-photon and Bob sends m-photon
\begin{equation}
  \begin{aligned}
    P_{nm}= e^{-r_1(\sin \theta_1+\cos \theta_1)} \frac{[r_1(\sin \theta_1+\cos \theta_1)]^n}{n !} \\ \times 
    e^{-r_2(\sin \theta_2+\cos \theta_2)} \frac{[r_2(\sin \theta_2+\cos \theta_2)]^m}{m !}.
  \end{aligned}
\end{equation}
$Y_{nm}$ is a function of $(\theta_1,\theta_2)$ only, independent of $(r_1,r_2)$, because the Fock state does not retain any information about the original pulse intensity. Our goal is to obtain a set of variables $Y_{nm}$ that are independent of the setup of the decoy state, thus enabling the construction of a linear programme. Therefore we need to perform a probabilistic post-selection step to shape the original intensity distribution, i.e. the users keep or discard the signals with probability $q_{\mu}=C\pi^2 \sqrt{\mu_H\left(\mu_{\max }-\mu_H\right) \mu_V\left(\mu_{\max }-\mu_V\right)}$.
Thus the actual intensity probability distribution becomes
\begin{equation}
  \begin{aligned}
    p_\mu^{\sim} = p_\mu q_\mu =C_1e^{(\mu_H+\mu_V)},
  \end{aligned}
\end{equation}
where $C_1$ is a normalization factor \cite{wangFullyPassiveQuantum2023}. Bob also get a new intensity probability distribution $p_{\mu^{\prime}}^{\sim}= C_2 e^{(\mu_H^{\prime}+\mu_V^{\prime})}$.

Combining the concentric sector post-selection regions $\left\{ S_i,S_j \right\}$ with the new intensity probability distribution $p_\mu^{\sim}$ and $p_{\mu^{\prime}}^{\sim}$, we will have
\begin{equation}
  \begin{aligned}
    &p_{\mu}^{\sim}(r_1, \theta_1) p_{\mu^{\prime}}^{\sim}(r_2, \theta_2) P_{nm}(r_{1}, r_{2}, \theta_{1}, \theta_{2})  \ r_1 r_2  \\
    \
    =&C_1\frac{r_1^{n+1}(\sin \theta_1+\cos\theta_1)^n}{n!}
    C_2\frac{r_2^{m+1}(\sin \theta_2+\cos\theta_2)^m}{m!}.
  \end{aligned}
\end{equation}
Note that the exponential term of the intensity distribution cancels the exponential term of the joint Poisson distribution, so it is possible to decouple the overall distribution in $r$ and $\theta$. So we have
\begin{widetext}
\begin{equation}
  \begin{aligned}
    \langle P_{nm}Y_{nm} \rangle _{S_iS_j}
    =&\frac{C_1C_2}{P_{S_iS_j}}\int_0^{r_1,{S_i}} \int_0^{r_2,{S_j}} 
    \frac{r_1^{n+1}}{n!} \frac{r_2^{m+1}}{m!} dr_1 dr_2  \\
     &\times
    \int_{\theta_1^{min}}^{\theta_1^{max}}  \int_{\theta_2^{min}}^{\theta_2^{max}} 
    (\sin \theta_1+\cos\theta_1)^n (\sin \theta_2+\cos\theta_2)^m  Y_{nm}(\theta_1,\theta_2)
     d\theta_1  d\theta_2.
    \label{eA4}
  \end{aligned}
\end{equation}
\end{widetext}
According to Eq.(\ref{eA1}), we can rewrite Eq.(\ref{eA4}) as
\begin{equation}
  \begin{aligned}
    \langle P_{nm}Y_{nm} \rangle_ {S_iS_j}= \langle P_{nm} \rangle _{S_iS_j} \times Y_{nm}^{\prime}, \\
  \end{aligned}
\end{equation}
similarly, the error yield also satisfies
\begin{equation}
  \begin{aligned}
     \langle P_{nm}E_{nm} \rangle_ {S_iS_j}= \langle P_{nm} \rangle _{S_iS_j} \times E_{nm}^{\prime}.
  \end{aligned}
\end{equation}
And the newly defined $Y_{nm}^{\prime}$ satisfies
\begin{widetext}
\begin{equation}
  \begin{aligned}
    Y_{nm}^{\prime}=\frac{\int_{\theta_1^{min}}^{\theta_1^{max}}  \int_{\theta_2^{min}}^{\theta_2^{max}} 
    (\sin \theta_1+\cos\theta_1)^n (\sin \theta_2+\cos\theta_2)^m  Y_{nm}(\theta_1,\theta_2)
     d\theta_1  d\theta_2}
     {\int_{\theta_1^{min}}^{\theta_1^{max}}  \int_{\theta_2^{min}}^{\theta_2^{max}} 
     (\sin \theta_1+\cos\theta_1)^n (\sin \theta_2+\cos\theta_2)^m
      d\theta_1  d\theta_2}.
  \end{aligned}
\end{equation}
\end{widetext}
Importantly, $Y_{nm}^{\prime}$ is the same for all decoy-state settings.
Thus taking $Y_{nm}^{\prime}$ as an independent variable, we can construct a linear program to further estimate $Y_{11}^{\mathrm{Z,L}}$ and $e_{11}^{\mathrm{X,U}}$. 

\section{\label{Appendix C}Numerical approaches}

To estimate $Y_{11}^{\mathrm{Z},L}$ and $e_{11}^{\mathrm{X,U}}$, we need to solve the following linear programming equations 
\begin{equation}
  \begin{aligned}
  & \min : Y_{11}^{\mathrm{Z}} \text {, } \\
  & \text { s.t. : } 0 \leqslant Y_{n m}^{\mathrm{Z}} \leqslant 1, n, m \in \mathcal{S}_{\text {cut }} \text {, } \\
  & \langle Q \rangle_{S_i^\mathrm{Z}S_j^\mathrm{Z}}  -\left(1-\sum_{n, m \in S_{\mathrm{cut}}} \langle P_{nm} \rangle_{S_i^\mathrm{Z}S_j^\mathrm{Z}} \right) \\
  & \leqslant \sum_{n, m \in \mathcal{S}_{\mathrm{cut}}} \langle P_{nm} \rangle_ {S_i^\mathrm{Z}S_j^\mathrm{Z}} Y_{n m}^{\mathrm{Z}} \leqslant \langle Q \rangle_{S_i^ZS_j^Z}. \\
  \\
  & \max : e_{11}^{\mathrm{X}} \text {, } \\
  & \text { s.t. : } 0 \leqslant Y_{n m}^{\mathrm{X}} \leqslant 1,0 \leqslant Y_{n m}^{\mathrm{X}} e_{n m}^{\mathrm{X}} \leqslant 1, n, m \in \mathcal{S}_{\text {cut }} \text {, } \\
  & \langle Q \rangle_{S_i^\mathrm{X}S_j^\mathrm{X}}  -\left(1-\sum_{n, m \in S_{\mathrm{cut}}} \langle P_{nm} \rangle_{S_i^\mathrm{X}S_j^\mathrm{X}} \right) \\
  & \leqslant \sum_{n, m \in \mathcal{S}_{\mathrm{cut}}} \langle P_{nm} \rangle_{S_i^\mathrm{X}S_j^\mathrm{X}} Y_{n m}^{\mathrm{X}} \leqslant \langle Q \rangle_{S_i^\mathrm{X}S_j^\mathrm{X}} \\
  & \langle T  \rangle_{S_i^\mathrm{X}S_j^\mathrm{X}} -\left(1-\sum_{n, m \in \mathcal{S}_{\mathrm{cut}}} \langle P_{nm} \rangle_{S_i^\mathrm{X}S_j^\mathrm{X}} \right) \\
  & \leqslant \sum_{n, m \in S_{\mathrm{cut}}} \langle P_{nm} \rangle_{S_i^\mathrm{X}S_j^\mathrm{X}}  e_{n m}^{\mathrm{X}} Y_{n m}^{\mathrm{X}}  \leqslant \langle T  \rangle_{S_i^\mathrm{X}S_j^\mathrm{X}}. \\
  &
  \label{e16}
  \end{aligned}
\end{equation}
where $\mathcal{S}_{cut}$ denotes a finite data set of $n$ and $m$, with $\mathcal{S}_{cut}=\left\{n, m \in \mathbb{N}\right.$ with $\left.n \leqslant N_{\text {cut }}, m \leqslant M_{\text {cut }}\right\}$. $N_{\text {cut }}$ and $M_{\text {cut }}$ are the threshold photon number for the decoy-state linear programs. $i$ and $j$ determine the choice of different decoy intensities.

\section{\label{Appendix B}MDI-QKD model}
In previous models \cite{ma2012alternative,xu2013practical,xu2014protocol,yu2013three,zhou2016making,jiang2021higher}, the total gain and QBER were calculated for the perfectly encoded BB84 states. However the passive scheme requires post-selection in a finite region for state preparation and therefore sends mixed states, so we provide a model that is able to obtain the gain and QBER for this imperfect state preparation case.
Assume that Alice and Bob prepare quantum states randomly with a passive source, we write as
$c_0 \ket{0} +c_1e^{i \phi}\ket{1}$ and $c_0^{\prime} \ket{0}+c_1^{\prime}e^{i \phi^{\prime}}\ket{1} $, the coefficient satisfies:
\begin{equation}
  \begin{aligned}
    c_0= \cos \frac{\theta}{2}, c_1=\sin \frac{\theta}{2}, \\
    c_0^{\prime}= \cos\frac{\theta^{\prime}}{2}, c_1^{\prime}=\sin \frac{\theta^{\prime}}{2},
  \end{aligned}
\end{equation}
where $\theta$ and $\theta^{\prime}$ are the polar angles, $\phi$ and $\phi^{\prime}$ are the azimuth angles on the Bloch sphere respectively.
So the relation between the coefficients $c_0$($c_0^{\prime}$) and $c_1$($c_1^{\prime}$) determines the specific form of the encoding states.
Alice (Bob) detect the specific form of the quantum state in each round and then use the intensity and phase information for post-selection, which determines the base vector (Z, X) and bits (0, 1) of the  state they sent:
\begin{equation}
  \begin{aligned}
    \ket{c_0\sqrt{\mu_A} e^{i \varphi}}_{A_0} \ket{c_1\sqrt{\mu_A} e^{i(\varphi+\phi)}}_{A_1}  \\
    \ket{c_0^{\prime}\sqrt{\mu_B} e^{i \varphi^{\prime}}}_{B_0} \ket{c_1^{\prime}\sqrt{\mu_B} e^{i (\varphi^{\prime}+\phi^{\prime})}}_{B_1},
  \end{aligned}
\end{equation}
where $\varphi$ and $\varphi^{\prime}$ are the overall randomized phases, $\mu_A$ and $\mu_B$ are the the intensity of the preparation, and the subscripts 0 and 1 of A (B) denote the horizontal polarization and vertical polarization, respectively. 

Transmitting through lossy channels, then passing through the beam splitters in the relay. With the overall efficiencies $\eta_A$ and $\eta_B$, the state arrived at the detectors can be transformed into four detection modes
\begin{equation}
  \begin{aligned}
    & \ket{\frac{c_0 \gamma_A e^{i \varphi}-c_0^{\prime} \gamma_B e^{i \varphi^{\prime}}}{\sqrt{2}}}_{D_{\rm{1H}}}
    \ket{\frac{c_0 \gamma_A e^{i \varphi}+c_0^{\prime} \gamma_B e^{i \varphi^{\prime}}}{\sqrt{2}}}_{D_{\rm{2H}}}   \\
    & \otimes 
    \ket{\frac{c_1 \gamma_A e^{i (\varphi+\phi)}-c_1^{\prime} \gamma_B e^{i (\varphi^{\prime}+\phi^{\prime})}}{\sqrt{2}}}_{D_{\rm{1V}}} \\
    & \otimes
    \ket{\frac{c_1 \gamma_A e^{i (\varphi+\phi)}+c_1^{\prime} \gamma_B e^{i (\varphi^{\prime}+\phi^{\prime})}}{\sqrt{2}}}_{D_{\rm{2V}}},
  \end{aligned}
\end{equation}
where $\gamma_A = \sqrt{\eta_A\mu_A}$, $\gamma_B = \sqrt{\eta_B\mu_B}$.
Then Charlie announces the results of the detector response results,
$\ket{\Psi^+}$ corresponds to coincident detections of $D_{1H} \& D_{1V}$ or $D_{2H} \& D_{2V}$, 
$\ket{\Psi^-}$ corresponds to coincident detections of $D_{1H} \& D_{2V}$ or $D_{1V} \& D_{2H}$, 
and others are invalid response.
For simplicity, we use the notations
\begin{equation}
  \begin{aligned}
  & \gamma=\frac{\eta_A\mu_A+\eta_B\mu_B}{2}, \\
  &\gamma_0=\frac{c_0^2 \eta_A\mu_A+c_0^{\prime 2}\eta_B\mu_B}{2}, 
  \gamma_1=\frac{c_1^2 \eta_A\mu_A+c_1^{\prime 2} \eta_B\mu_B}{2} \\
  & \beta_0=c_0 c_0^{\prime} \sqrt{\eta_A\mu_A\eta_B\mu_B}, \beta_1=c_1 c_1^{\prime} \sqrt{\eta_A\mu_A\eta_B\mu_B}
  \end{aligned}
\end{equation}
Thus, for different combinations of intensity and encoding sent by Alice and Bob, we can obtain the gain when the detection result is $\ket{\Psi^+}$ or $\ket{\Psi^-}$. For a specific intensity combination, we define the gain $Q_{\Psi^{-}}^{\Omega \ S}$ and $Q_{\Psi^{+}}^{\Omega\ S}$, $\Omega \in \left\{\rm{Z},\rm{X}\right\}$ represents the basis selection, $S \in \left\{\rm{HH,HV,VH,VV,++,+-,-+,--}\right\}$ represents the combination of encoding states when Alice and Bob select the same basis. After integrating over $[0,2\pi)$ for the difference between random phases $\varphi$ and $\varphi^{\prime}$, we have
\begin{widetext}
\begin{equation}
  \begin{aligned}
    Q_{\Psi^{-}}^{\Omega \ S} & =2(1-p_d)^2 e^{-\gamma}\left\{1+\frac{1}{4}(\beta_0^2+\beta_1^2-2\beta_0\beta_1\cos\phi)+\left(1-p_d\right)^2 e^{-\gamma}
   -\left(1-p_d\right)\left[e^{-\gamma_0} I_0(\beta_1)+e^{-\gamma_1} I_0(\beta_0)\right]\right\}, \\
    Q_{\Psi^{+}}^{\Omega \ S} & =2(1-p_d)^2 e^{-\gamma}\left\{1+\frac{1}{4}(\beta_0^2+\beta_1^2+2\beta_0\beta_1\cos\phi)  +(1-p_d)^2 e^{-\gamma}
   -\left(1-p_d\right)\left[e^{-\gamma_0} I_0(\beta_1)+e^{-\gamma_1} I_0(\beta_0)\right]\right\}.
  \end{aligned}
\end{equation}
\end{widetext} 
$I_0(x)$ is the modified Bessel function of the first kind. For a small value of $x$, we can take the first-order approximation $I_0(x) \approx 1+x^2/4$. $\phi=\phi - \phi^{\prime}$ is the difference between the azimuth angles.

Moreover, the QBER can be obtained by the gian $Q_{\Psi^{-}}^{\Omega \ S}$ and $Q_{\Psi^{+}}^{\Omega \ S}$
\begin{equation}
  \begin{aligned}
    e^{\Omega}_{\Gamma}=e_d\left(1-\hat{e}_{\Gamma}^{\Omega}\right)+\left(1-e_d\right) \hat{e}_{\Gamma}^{\Omega},
  \end{aligned}
\end{equation}
where $e_d$ is the misalignment-error probability, $\Omega \in \left\{\rm{Z},\rm{X}\right\}$, $\Gamma \in \left\{\Psi^{-},\Psi^{+}\right\}$. And $\hat{e}_{\Gamma}^{\Omega}$ is the QBER without $e_d$, which satisfies
\begin{equation}
  \begin{aligned}
    \hat{e}_{\Psi^{-}}^{\rm{Z}}=\frac{Q_{\Psi^{-}}^{{\rm{Z}}, \rm{HH}}+Q_{\Psi^{-}}^{\rm{Z}, \rm{VV}}}{Q_{\Psi^{-}}^{\rm{Z},\rm{HH}}+Q_{\Psi^{-}}^{\rm{Z},\rm{HV}}+Q_{\Psi^{-}}^{\rm{Z},\rm{VH}}+Q_{\Psi^{-}}^{\rm{Z}, \rm{VV}}},   \\ 
    \hat{e}_{\Psi^{-}}^{\rm{X}}=\frac{Q_{\Psi^{-}}^{{\rm{X}}, ++}+Q_{\Psi^{-}}^{\rm{X}, --}}{Q_{\Psi^{-}}^{\rm{X},++}+Q_{\Psi^{-}}^{\rm{X},+-}+Q_{\Psi^{-}}^{\rm{X},-+}+Q_{\Psi^{-}}^{\rm{X}, --}}. 
  \end{aligned}
\end{equation}

Therefore, even if the users cannot prepare the BB84 states perfectly, as long as they know the exact form of the state in the current turn, our model can always calculate the corresponding gain and QBER.

\bibliography{apssamp.bib}

\begin{thebibliography}{51}%
\makeatletter
\providecommand \@ifxundefined [1]{%
 \@ifx{#1\undefined}
}%
\providecommand \@ifnum [1]{%
 \ifnum #1\expandafter \@firstoftwo
 \else \expandafter \@secondoftwo
 \fi
}%
\providecommand \@ifx [1]{%
 \ifx #1\expandafter \@firstoftwo
 \else \expandafter \@secondoftwo
 \fi
}%
\providecommand \natexlab [1]{#1}%
\providecommand \enquote  [1]{``#1''}%
\providecommand \bibnamefont  [1]{#1}%
\providecommand \bibfnamefont [1]{#1}%
\providecommand \citenamefont [1]{#1}%
\providecommand \href@noop [0]{\@secondoftwo}%
\providecommand \href [0]{\begingroup \@sanitize@url \@href}%
\providecommand \@href[1]{\@@startlink{#1}\@@href}%
\providecommand \@@href[1]{\endgroup#1\@@endlink}%
\providecommand \@sanitize@url [0]{\catcode `\\12\catcode `\$12\catcode
  `\&12\catcode `\#12\catcode `\^12\catcode `\_12\catcode `\%12\relax}%
\providecommand \@@startlink[1]{}%
\providecommand \@@endlink[0]{}%
\providecommand \url  [0]{\begingroup\@sanitize@url \@url }%
\providecommand \@url [1]{\endgroup\@href {#1}{\urlprefix }}%
\providecommand \urlprefix  [0]{URL }%
\providecommand \Eprint [0]{\href }%
\providecommand \doibase [0]{https://doi.org/}%
\providecommand \selectlanguage [0]{\@gobble}%
\providecommand \bibinfo  [0]{\@secondoftwo}%
\providecommand \bibfield  [0]{\@secondoftwo}%
\providecommand \translation [1]{[#1]}%
\providecommand \BibitemOpen [0]{}%
\providecommand \bibitemStop [0]{}%
\providecommand \bibitemNoStop [0]{.\EOS\space}%
\providecommand \EOS [0]{\spacefactor3000\relax}%
\providecommand \BibitemShut  [1]{\csname bibitem#1\endcsname}%
\let\auto@bib@innerbib\@empty
\bibitem [{\citenamefont {Bennett}\ \emph {et~al.}(1984)\citenamefont
  {Bennett}, \citenamefont {Brassard} \emph {et~al.}}]{bennett1984proceedings}%
  \BibitemOpen
  \bibfield  {author} {\bibinfo {author} {\bibfnamefont {C.~H.}\ \bibnamefont
  {Bennett}}, \bibinfo {author} {\bibfnamefont {G.}~\bibnamefont {Brassard}},
  \emph {et~al.},\ }\href@noop {} {\bibinfo {title} {Proceedings of the ieee
  international conference on computers, systems and signal processing}}
  (\bibinfo {year} {1984})\BibitemShut {NoStop}%
\bibitem [{\citenamefont {Ekert}(1991)}]{ekert1991quantum}%
  \BibitemOpen
  \bibfield  {author} {\bibinfo {author} {\bibfnamefont {A.~K.}\ \bibnamefont
  {Ekert}},\ }\href {https://doi.org/10.1103/PhysRevLett.67.661} {\bibfield
  {journal} {\bibinfo  {journal} {Physical Review Letters}\ }\textbf {\bibinfo
  {volume} {67}},\ \bibinfo {pages} {661} (\bibinfo {year} {1991})}\BibitemShut
  {NoStop}%
\bibitem [{\citenamefont {Lo}\ and\ \citenamefont
  {Chau}(1999)}]{lo1999unconditional}%
  \BibitemOpen
  \bibfield  {author} {\bibinfo {author} {\bibfnamefont {H.-K.}\ \bibnamefont
  {Lo}}\ and\ \bibinfo {author} {\bibfnamefont {H.~F.}\ \bibnamefont {Chau}},\
  }\href {https://doi.org/10.1126/science.283.5410.2050} {\bibfield  {journal}
  {\bibinfo  {journal} {Science}\ }\textbf {\bibinfo {volume} {283}},\ \bibinfo
  {pages} {2050} (\bibinfo {year} {1999})}\BibitemShut {NoStop}%
\bibitem [{\citenamefont {Shor}\ and\ \citenamefont
  {Preskill}(2000)}]{shor2000simple}%
  \BibitemOpen
  \bibfield  {author} {\bibinfo {author} {\bibfnamefont {P.~W.}\ \bibnamefont
  {Shor}}\ and\ \bibinfo {author} {\bibfnamefont {J.}~\bibnamefont
  {Preskill}},\ }\href {https://doi.org/10.1103/PhysRevLett.85.441} {\bibfield
  {journal} {\bibinfo  {journal} {Physical Review Letters}\ }\textbf {\bibinfo
  {volume} {85}},\ \bibinfo {pages} {441} (\bibinfo {year} {2000})}\BibitemShut
  {NoStop}%
\bibitem [{\citenamefont {Renner}(2008)}]{renner2008security}%
  \BibitemOpen
  \bibfield  {author} {\bibinfo {author} {\bibfnamefont {R.}~\bibnamefont
  {Renner}},\ }\href {https://doi.org/10.1142/S0219749908003256} {\bibfield
  {journal} {\bibinfo  {journal} {International Journal of Quantum
  Information}\ }\textbf {\bibinfo {volume} {6}},\ \bibinfo {pages} {1}
  (\bibinfo {year} {2008})}\BibitemShut {NoStop}%
\bibitem [{\citenamefont {Portmann}\ and\ \citenamefont
  {Renner}(2022)}]{portmann2022security}%
  \BibitemOpen
  \bibfield  {author} {\bibinfo {author} {\bibfnamefont {C.}~\bibnamefont
  {Portmann}}\ and\ \bibinfo {author} {\bibfnamefont {R.}~\bibnamefont
  {Renner}},\ }\href {https://doi.org/10.1103/RevModPhys.94.025008} {\bibfield
  {journal} {\bibinfo  {journal} {Reviews of Modern Physics}\ }\textbf
  {\bibinfo {volume} {94}},\ \bibinfo {pages} {025008} (\bibinfo {year}
  {2022})}\BibitemShut {NoStop}%
\bibitem [{\citenamefont {Scarani}\ \emph {et~al.}(2009)\citenamefont
  {Scarani}, \citenamefont {Bechmann-Pasquinucci}, \citenamefont {Cerf},
  \citenamefont {Du{\v{s}}ek}, \citenamefont {L{\"u}tkenhaus},\ and\
  \citenamefont {Peev}}]{scarani2009security}%
  \BibitemOpen
  \bibfield  {author} {\bibinfo {author} {\bibfnamefont {V.}~\bibnamefont
  {Scarani}}, \bibinfo {author} {\bibfnamefont {H.}~\bibnamefont
  {Bechmann-Pasquinucci}}, \bibinfo {author} {\bibfnamefont {N.~J.}\
  \bibnamefont {Cerf}}, \bibinfo {author} {\bibfnamefont {M.}~\bibnamefont
  {Du{\v{s}}ek}}, \bibinfo {author} {\bibfnamefont {N.}~\bibnamefont
  {L{\"u}tkenhaus}},\ and\ \bibinfo {author} {\bibfnamefont {M.}~\bibnamefont
  {Peev}},\ }\href {https://doi.org/10.1103/RevModPhys.81.1301} {\bibfield
  {journal} {\bibinfo  {journal} {Reviews of Modern Physics}\ }\textbf
  {\bibinfo {volume} {81}},\ \bibinfo {pages} {1301} (\bibinfo {year}
  {2009})}\BibitemShut {NoStop}%
\bibitem [{\citenamefont {Jain}\ \emph {et~al.}(2016)\citenamefont {Jain},
  \citenamefont {Stiller}, \citenamefont {Khan}, \citenamefont {Elser},
  \citenamefont {Marquardt},\ and\ \citenamefont {Leuchs}}]{jain2016attacks}%
  \BibitemOpen
  \bibfield  {author} {\bibinfo {author} {\bibfnamefont {N.}~\bibnamefont
  {Jain}}, \bibinfo {author} {\bibfnamefont {B.}~\bibnamefont {Stiller}},
  \bibinfo {author} {\bibfnamefont {I.}~\bibnamefont {Khan}}, \bibinfo {author}
  {\bibfnamefont {D.}~\bibnamefont {Elser}}, \bibinfo {author} {\bibfnamefont
  {C.}~\bibnamefont {Marquardt}},\ and\ \bibinfo {author} {\bibfnamefont
  {G.}~\bibnamefont {Leuchs}},\ }\href
  {https://doi.org/10.1080/00107514.2016.1148333} {\bibfield  {journal}
  {\bibinfo  {journal} {Contemporary Physics}\ }\textbf {\bibinfo {volume}
  {57}},\ \bibinfo {pages} {366} (\bibinfo {year} {2016})}\BibitemShut
  {NoStop}%
\bibitem [{\citenamefont {Xu}\ \emph {et~al.}(2020)\citenamefont {Xu},
  \citenamefont {Ma}, \citenamefont {Zhang}, \citenamefont {Lo},\ and\
  \citenamefont {Pan}}]{xu2020secure}%
  \BibitemOpen
  \bibfield  {author} {\bibinfo {author} {\bibfnamefont {F.}~\bibnamefont
  {Xu}}, \bibinfo {author} {\bibfnamefont {X.}~\bibnamefont {Ma}}, \bibinfo
  {author} {\bibfnamefont {Q.}~\bibnamefont {Zhang}}, \bibinfo {author}
  {\bibfnamefont {H.-K.}\ \bibnamefont {Lo}},\ and\ \bibinfo {author}
  {\bibfnamefont {J.-W.}\ \bibnamefont {Pan}},\ }\href
  {https://doi.org/10.1103/RevModPhys.92.025002} {\bibfield  {journal}
  {\bibinfo  {journal} {Reviews of Modern Physics}\ }\textbf {\bibinfo {volume}
  {92}},\ \bibinfo {pages} {025002} (\bibinfo {year} {2020})}\BibitemShut
  {NoStop}%
\bibitem [{\citenamefont {Lo}\ \emph {et~al.}(2012)\citenamefont {Lo},
  \citenamefont {Curty},\ and\ \citenamefont {Qi}}]{lo2012measurement}%
  \BibitemOpen
  \bibfield  {author} {\bibinfo {author} {\bibfnamefont {H.-K.}\ \bibnamefont
  {Lo}}, \bibinfo {author} {\bibfnamefont {M.}~\bibnamefont {Curty}},\ and\
  \bibinfo {author} {\bibfnamefont {B.}~\bibnamefont {Qi}},\ }\href
  {https://doi.org/10.1103/PhysRevLett.108.130503} {\bibfield  {journal}
  {\bibinfo  {journal} {Physical Review Letters}\ }\textbf {\bibinfo {volume}
  {108}},\ \bibinfo {pages} {130503} (\bibinfo {year} {2012})}\BibitemShut
  {NoStop}%
\bibitem [{\citenamefont {Braunstein}\ and\ \citenamefont
  {Pirandola}(2012)}]{braunstein2012side}%
  \BibitemOpen
  \bibfield  {author} {\bibinfo {author} {\bibfnamefont {S.~L.}\ \bibnamefont
  {Braunstein}}\ and\ \bibinfo {author} {\bibfnamefont {S.}~\bibnamefont
  {Pirandola}},\ }\href {https://doi.org/10.1103/PhysRevLett.108.130502}
  {\bibfield  {journal} {\bibinfo  {journal} {Physical Review Letters}\
  }\textbf {\bibinfo {volume} {108}},\ \bibinfo {pages} {130502} (\bibinfo
  {year} {2012})}\BibitemShut {NoStop}%
\bibitem [{\citenamefont {Tang}\ \emph {et~al.}(2014)\citenamefont {Tang},
  \citenamefont {Yin}, \citenamefont {Chen}, \citenamefont {Liu}, \citenamefont
  {Zhang}, \citenamefont {Jiang}, \citenamefont {Zhang}, \citenamefont {Wang},
  \citenamefont {You}, \citenamefont {Guan} \emph
  {et~al.}}]{tang2014measurement}%
  \BibitemOpen
  \bibfield  {author} {\bibinfo {author} {\bibfnamefont {Y.-L.}\ \bibnamefont
  {Tang}}, \bibinfo {author} {\bibfnamefont {H.-L.}\ \bibnamefont {Yin}},
  \bibinfo {author} {\bibfnamefont {S.-J.}\ \bibnamefont {Chen}}, \bibinfo
  {author} {\bibfnamefont {Y.}~\bibnamefont {Liu}}, \bibinfo {author}
  {\bibfnamefont {W.-J.}\ \bibnamefont {Zhang}}, \bibinfo {author}
  {\bibfnamefont {X.}~\bibnamefont {Jiang}}, \bibinfo {author} {\bibfnamefont
  {L.}~\bibnamefont {Zhang}}, \bibinfo {author} {\bibfnamefont
  {J.}~\bibnamefont {Wang}}, \bibinfo {author} {\bibfnamefont {L.-X.}\
  \bibnamefont {You}}, \bibinfo {author} {\bibfnamefont {J.-Y.}\ \bibnamefont
  {Guan}}, \emph {et~al.},\ }\href
  {https://doi.org/10.1103/PhysRevLett.113.190501} {\bibfield  {journal}
  {\bibinfo  {journal} {Physical Review Letters}\ }\textbf {\bibinfo {volume}
  {113}},\ \bibinfo {pages} {190501} (\bibinfo {year} {2014})}\BibitemShut
  {NoStop}%
\bibitem [{\citenamefont {Wang}\ \emph {et~al.}(2015)\citenamefont {Wang},
  \citenamefont {Song}, \citenamefont {Yin}, \citenamefont {Wang},
  \citenamefont {Chen}, \citenamefont {Zhang}, \citenamefont {Guo},\ and\
  \citenamefont {Han}}]{wang2015phase}%
  \BibitemOpen
  \bibfield  {author} {\bibinfo {author} {\bibfnamefont {C.}~\bibnamefont
  {Wang}}, \bibinfo {author} {\bibfnamefont {X.-T.}\ \bibnamefont {Song}},
  \bibinfo {author} {\bibfnamefont {Z.-Q.}\ \bibnamefont {Yin}}, \bibinfo
  {author} {\bibfnamefont {S.}~\bibnamefont {Wang}}, \bibinfo {author}
  {\bibfnamefont {W.}~\bibnamefont {Chen}}, \bibinfo {author} {\bibfnamefont
  {C.-M.}\ \bibnamefont {Zhang}}, \bibinfo {author} {\bibfnamefont {G.-C.}\
  \bibnamefont {Guo}},\ and\ \bibinfo {author} {\bibfnamefont {Z.-F.}\
  \bibnamefont {Han}},\ }\href {https://doi.org/10.1103/PhysRevLett.115.160502}
  {\bibfield  {journal} {\bibinfo  {journal} {Physical Review Letters}\
  }\textbf {\bibinfo {volume} {115}},\ \bibinfo {pages} {160502} (\bibinfo
  {year} {2015})}\BibitemShut {NoStop}%
\bibitem [{\citenamefont {Yin}\ \emph {et~al.}(2016)\citenamefont {Yin},
  \citenamefont {Chen}, \citenamefont {Yu}, \citenamefont {Liu}, \citenamefont
  {You}, \citenamefont {Zhou}, \citenamefont {Chen}, \citenamefont {Mao},
  \citenamefont {Huang}, \citenamefont {Zhang} \emph
  {et~al.}}]{yin2016measurement}%
  \BibitemOpen
  \bibfield  {author} {\bibinfo {author} {\bibfnamefont {H.-L.}\ \bibnamefont
  {Yin}}, \bibinfo {author} {\bibfnamefont {T.-Y.}\ \bibnamefont {Chen}},
  \bibinfo {author} {\bibfnamefont {Z.-W.}\ \bibnamefont {Yu}}, \bibinfo
  {author} {\bibfnamefont {H.}~\bibnamefont {Liu}}, \bibinfo {author}
  {\bibfnamefont {L.-X.}\ \bibnamefont {You}}, \bibinfo {author} {\bibfnamefont
  {Y.-H.}\ \bibnamefont {Zhou}}, \bibinfo {author} {\bibfnamefont {S.-J.}\
  \bibnamefont {Chen}}, \bibinfo {author} {\bibfnamefont {Y.}~\bibnamefont
  {Mao}}, \bibinfo {author} {\bibfnamefont {M.-Q.}\ \bibnamefont {Huang}},
  \bibinfo {author} {\bibfnamefont {W.-J.}\ \bibnamefont {Zhang}}, \emph
  {et~al.},\ }\href {https://doi.org/10.1103/PhysRevLett.117.190501} {\bibfield
   {journal} {\bibinfo  {journal} {Physical Review Letters}\ }\textbf {\bibinfo
  {volume} {117}},\ \bibinfo {pages} {190501} (\bibinfo {year}
  {2016})}\BibitemShut {NoStop}%
\bibitem [{\citenamefont {Pirandola}\ \emph {et~al.}(2015)\citenamefont
  {Pirandola}, \citenamefont {Ottaviani}, \citenamefont {Spedalieri},
  \citenamefont {Weedbrook}, \citenamefont {Braunstein}, \citenamefont {Lloyd},
  \citenamefont {Gehring}, \citenamefont {Jacobsen},\ and\ \citenamefont
  {Andersen}}]{pirandola2015high}%
  \BibitemOpen
  \bibfield  {author} {\bibinfo {author} {\bibfnamefont {S.}~\bibnamefont
  {Pirandola}}, \bibinfo {author} {\bibfnamefont {C.}~\bibnamefont
  {Ottaviani}}, \bibinfo {author} {\bibfnamefont {G.}~\bibnamefont
  {Spedalieri}}, \bibinfo {author} {\bibfnamefont {C.}~\bibnamefont
  {Weedbrook}}, \bibinfo {author} {\bibfnamefont {S.~L.}\ \bibnamefont
  {Braunstein}}, \bibinfo {author} {\bibfnamefont {S.}~\bibnamefont {Lloyd}},
  \bibinfo {author} {\bibfnamefont {T.}~\bibnamefont {Gehring}}, \bibinfo
  {author} {\bibfnamefont {C.~S.}\ \bibnamefont {Jacobsen}},\ and\ \bibinfo
  {author} {\bibfnamefont {U.~L.}\ \bibnamefont {Andersen}},\ }\href
  {https://doi.org/10.1038/nphoton.2015.83} {\bibfield  {journal} {\bibinfo
  {journal} {Nature Photonics}\ }\textbf {\bibinfo {volume} {9}},\ \bibinfo
  {pages} {397} (\bibinfo {year} {2015})}\BibitemShut {NoStop}%
\bibitem [{\citenamefont {Liu}\ \emph {et~al.}(2019)\citenamefont {Liu},
  \citenamefont {Wang}, \citenamefont {Wei}, \citenamefont {Fang},
  \citenamefont {Li}, \citenamefont {Liu}, \citenamefont {Liang}, \citenamefont
  {Zhang}, \citenamefont {Zhang}, \citenamefont {Li} \emph
  {et~al.}}]{liu2019experimental}%
  \BibitemOpen
  \bibfield  {author} {\bibinfo {author} {\bibfnamefont {H.}~\bibnamefont
  {Liu}}, \bibinfo {author} {\bibfnamefont {W.}~\bibnamefont {Wang}}, \bibinfo
  {author} {\bibfnamefont {K.}~\bibnamefont {Wei}}, \bibinfo {author}
  {\bibfnamefont {X.-T.}\ \bibnamefont {Fang}}, \bibinfo {author}
  {\bibfnamefont {L.}~\bibnamefont {Li}}, \bibinfo {author} {\bibfnamefont
  {N.-L.}\ \bibnamefont {Liu}}, \bibinfo {author} {\bibfnamefont
  {H.}~\bibnamefont {Liang}}, \bibinfo {author} {\bibfnamefont {S.-J.}\
  \bibnamefont {Zhang}}, \bibinfo {author} {\bibfnamefont {W.}~\bibnamefont
  {Zhang}}, \bibinfo {author} {\bibfnamefont {H.}~\bibnamefont {Li}}, \emph
  {et~al.},\ }\href {https://doi.org/10.1103/PhysRevLett.122.160501} {\bibfield
   {journal} {\bibinfo  {journal} {Physical Review Letters}\ }\textbf {\bibinfo
  {volume} {122}},\ \bibinfo {pages} {160501} (\bibinfo {year}
  {2019})}\BibitemShut {NoStop}%
\bibitem [{\citenamefont {Semenenko}\ \emph {et~al.}(2020)\citenamefont
  {Semenenko}, \citenamefont {Sibson}, \citenamefont {Hart}, \citenamefont
  {Thompson}, \citenamefont {Rarity},\ and\ \citenamefont
  {Erven}}]{semenenko2020chip}%
  \BibitemOpen
  \bibfield  {author} {\bibinfo {author} {\bibfnamefont {H.}~\bibnamefont
  {Semenenko}}, \bibinfo {author} {\bibfnamefont {P.}~\bibnamefont {Sibson}},
  \bibinfo {author} {\bibfnamefont {A.}~\bibnamefont {Hart}}, \bibinfo {author}
  {\bibfnamefont {M.~G.}\ \bibnamefont {Thompson}}, \bibinfo {author}
  {\bibfnamefont {J.~G.}\ \bibnamefont {Rarity}},\ and\ \bibinfo {author}
  {\bibfnamefont {C.}~\bibnamefont {Erven}},\ }\href
  {https://doi.org/10.1364/OPTICA.379679} {\bibfield  {journal} {\bibinfo
  {journal} {Optica}\ }\textbf {\bibinfo {volume} {7}},\ \bibinfo {pages} {238}
  (\bibinfo {year} {2020})}\BibitemShut {NoStop}%
\bibitem [{\citenamefont {Wei}\ \emph {et~al.}(2020)\citenamefont {Wei},
  \citenamefont {Li}, \citenamefont {Tan}, \citenamefont {Li}, \citenamefont
  {Min}, \citenamefont {Zhang}, \citenamefont {Li}, \citenamefont {You},
  \citenamefont {Wang}, \citenamefont {Jiang} \emph {et~al.}}]{wei2020high}%
  \BibitemOpen
  \bibfield  {author} {\bibinfo {author} {\bibfnamefont {K.}~\bibnamefont
  {Wei}}, \bibinfo {author} {\bibfnamefont {W.}~\bibnamefont {Li}}, \bibinfo
  {author} {\bibfnamefont {H.}~\bibnamefont {Tan}}, \bibinfo {author}
  {\bibfnamefont {Y.}~\bibnamefont {Li}}, \bibinfo {author} {\bibfnamefont
  {H.}~\bibnamefont {Min}}, \bibinfo {author} {\bibfnamefont {W.-J.}\
  \bibnamefont {Zhang}}, \bibinfo {author} {\bibfnamefont {H.}~\bibnamefont
  {Li}}, \bibinfo {author} {\bibfnamefont {L.}~\bibnamefont {You}}, \bibinfo
  {author} {\bibfnamefont {Z.}~\bibnamefont {Wang}}, \bibinfo {author}
  {\bibfnamefont {X.}~\bibnamefont {Jiang}}, \emph {et~al.},\ }\href
  {https://doi.org/10.1103/PhysRevX.10.031030} {\bibfield  {journal} {\bibinfo
  {journal} {Physical Review X}\ }\textbf {\bibinfo {volume} {10}},\ \bibinfo
  {pages} {031030} (\bibinfo {year} {2020})}\BibitemShut {NoStop}%
\bibitem [{\citenamefont {Comandar}\ \emph {et~al.}(2016)\citenamefont
  {Comandar}, \citenamefont {Lucamarini}, \citenamefont {Fr{\"o}hlich},
  \citenamefont {Dynes}, \citenamefont {Sharpe}, \citenamefont {Tam},
  \citenamefont {Yuan}, \citenamefont {Penty},\ and\ \citenamefont
  {Shields}}]{comandar2016quantum}%
  \BibitemOpen
  \bibfield  {author} {\bibinfo {author} {\bibfnamefont {L.}~\bibnamefont
  {Comandar}}, \bibinfo {author} {\bibfnamefont {M.}~\bibnamefont
  {Lucamarini}}, \bibinfo {author} {\bibfnamefont {B.}~\bibnamefont
  {Fr{\"o}hlich}}, \bibinfo {author} {\bibfnamefont {J.}~\bibnamefont {Dynes}},
  \bibinfo {author} {\bibfnamefont {A.}~\bibnamefont {Sharpe}}, \bibinfo
  {author} {\bibfnamefont {S.-B.}\ \bibnamefont {Tam}}, \bibinfo {author}
  {\bibfnamefont {Z.}~\bibnamefont {Yuan}}, \bibinfo {author} {\bibfnamefont
  {R.}~\bibnamefont {Penty}},\ and\ \bibinfo {author} {\bibfnamefont
  {A.}~\bibnamefont {Shields}},\ }\href
  {https://doi.org/10.1038/nphoton.2016.50} {\bibfield  {journal} {\bibinfo
  {journal} {Nature Photonics}\ }\textbf {\bibinfo {volume} {10}},\ \bibinfo
  {pages} {312} (\bibinfo {year} {2016})}\BibitemShut {NoStop}%
\bibitem [{\citenamefont {Woodward}\ \emph {et~al.}(2021)\citenamefont
  {Woodward}, \citenamefont {Lo}, \citenamefont {Pittaluga}, \citenamefont
  {Minder}, \citenamefont {Para{\"\i}so}, \citenamefont {Lucamarini},
  \citenamefont {Yuan},\ and\ \citenamefont {Shields}}]{woodward2021gigahertz}%
  \BibitemOpen
  \bibfield  {author} {\bibinfo {author} {\bibfnamefont {R.~I.}\ \bibnamefont
  {Woodward}}, \bibinfo {author} {\bibfnamefont {Y.}~\bibnamefont {Lo}},
  \bibinfo {author} {\bibfnamefont {M.}~\bibnamefont {Pittaluga}}, \bibinfo
  {author} {\bibfnamefont {M.}~\bibnamefont {Minder}}, \bibinfo {author}
  {\bibfnamefont {T.}~\bibnamefont {Para{\"\i}so}}, \bibinfo {author}
  {\bibfnamefont {M.}~\bibnamefont {Lucamarini}}, \bibinfo {author}
  {\bibfnamefont {Z.}~\bibnamefont {Yuan}},\ and\ \bibinfo {author}
  {\bibfnamefont {A.}~\bibnamefont {Shields}},\ }\href
  {https://doi.org/10.1038/s41534-021-00394-2} {\bibfield  {journal} {\bibinfo
  {journal} {npj Quantum Information}\ }\textbf {\bibinfo {volume} {7}},\
  \bibinfo {pages} {58} (\bibinfo {year} {2021})}\BibitemShut {NoStop}%
\bibitem [{\citenamefont {Fan-Yuan}\ \emph {et~al.}(2022)\citenamefont
  {Fan-Yuan}, \citenamefont {Lu}, \citenamefont {Wang}, \citenamefont {Yin},
  \citenamefont {He}, \citenamefont {Chen}, \citenamefont {Zhou}, \citenamefont
  {Wang}, \citenamefont {Teng}, \citenamefont {Guo} \emph
  {et~al.}}]{fan2022robust}%
  \BibitemOpen
  \bibfield  {author} {\bibinfo {author} {\bibfnamefont {G.-J.}\ \bibnamefont
  {Fan-Yuan}}, \bibinfo {author} {\bibfnamefont {F.-Y.}\ \bibnamefont {Lu}},
  \bibinfo {author} {\bibfnamefont {S.}~\bibnamefont {Wang}}, \bibinfo {author}
  {\bibfnamefont {Z.-Q.}\ \bibnamefont {Yin}}, \bibinfo {author} {\bibfnamefont
  {D.-Y.}\ \bibnamefont {He}}, \bibinfo {author} {\bibfnamefont
  {W.}~\bibnamefont {Chen}}, \bibinfo {author} {\bibfnamefont {Z.}~\bibnamefont
  {Zhou}}, \bibinfo {author} {\bibfnamefont {Z.-H.}\ \bibnamefont {Wang}},
  \bibinfo {author} {\bibfnamefont {J.}~\bibnamefont {Teng}}, \bibinfo {author}
  {\bibfnamefont {G.-C.}\ \bibnamefont {Guo}}, \emph {et~al.},\ }\href
  {https://doi.org/10.1364/OPTICA.458937} {\bibfield  {journal} {\bibinfo
  {journal} {Optica}\ }\textbf {\bibinfo {volume} {9}},\ \bibinfo {pages} {812}
  (\bibinfo {year} {2022})}\BibitemShut {NoStop}%
\bibitem [{\citenamefont {Lu}\ \emph {et~al.}(2022)\citenamefont {Lu},
  \citenamefont {Wang}, \citenamefont {Yin}, \citenamefont {Wang},
  \citenamefont {Wang}, \citenamefont {Fan-Yuan}, \citenamefont {Huang},
  \citenamefont {He}, \citenamefont {Chen}, \citenamefont {Zhou} \emph
  {et~al.}}]{lu2022unbalanced}%
  \BibitemOpen
  \bibfield  {author} {\bibinfo {author} {\bibfnamefont {F.-Y.}\ \bibnamefont
  {Lu}}, \bibinfo {author} {\bibfnamefont {Z.-H.}\ \bibnamefont {Wang}},
  \bibinfo {author} {\bibfnamefont {Z.-Q.}\ \bibnamefont {Yin}}, \bibinfo
  {author} {\bibfnamefont {S.}~\bibnamefont {Wang}}, \bibinfo {author}
  {\bibfnamefont {R.}~\bibnamefont {Wang}}, \bibinfo {author} {\bibfnamefont
  {G.-J.}\ \bibnamefont {Fan-Yuan}}, \bibinfo {author} {\bibfnamefont {X.-J.}\
  \bibnamefont {Huang}}, \bibinfo {author} {\bibfnamefont {D.-Y.}\ \bibnamefont
  {He}}, \bibinfo {author} {\bibfnamefont {W.}~\bibnamefont {Chen}}, \bibinfo
  {author} {\bibfnamefont {Z.}~\bibnamefont {Zhou}}, \emph {et~al.},\ }\href
  {https://doi.org/10.1364/OPTICA.454228} {\bibfield  {journal} {\bibinfo
  {journal} {Optica}\ }\textbf {\bibinfo {volume} {9}},\ \bibinfo {pages} {886}
  (\bibinfo {year} {2022})}\BibitemShut {NoStop}%
\bibitem [{\citenamefont {Gisin}\ \emph {et~al.}(2006)\citenamefont {Gisin},
  \citenamefont {Fasel}, \citenamefont {Kraus}, \citenamefont {Zbinden},\ and\
  \citenamefont {Ribordy}}]{gisinTrojanhorseAttacksQuantumkeydistribution2006}%
  \BibitemOpen
  \bibfield  {author} {\bibinfo {author} {\bibfnamefont {N.}~\bibnamefont
  {Gisin}}, \bibinfo {author} {\bibfnamefont {S.}~\bibnamefont {Fasel}},
  \bibinfo {author} {\bibfnamefont {B.}~\bibnamefont {Kraus}}, \bibinfo
  {author} {\bibfnamefont {H.}~\bibnamefont {Zbinden}},\ and\ \bibinfo {author}
  {\bibfnamefont {G.}~\bibnamefont {Ribordy}},\ }\href
  {https://doi.org/10.1103/PhysRevA.73.022320} {\bibfield  {journal} {\bibinfo
  {journal} {Physical Review A}\ }\textbf {\bibinfo {volume} {73}},\ \bibinfo
  {pages} {022320} (\bibinfo {year} {2006})}\BibitemShut {NoStop}%
\bibitem [{\citenamefont {Jain}\ \emph
  {et~al.}(2014{\natexlab{a}})\citenamefont {Jain}, \citenamefont {Stiller},
  \citenamefont {Khan}, \citenamefont {Makarov}, \citenamefont {Marquardt},\
  and\ \citenamefont {Leuchs}}]{jain2014risk}%
  \BibitemOpen
  \bibfield  {author} {\bibinfo {author} {\bibfnamefont {N.}~\bibnamefont
  {Jain}}, \bibinfo {author} {\bibfnamefont {B.}~\bibnamefont {Stiller}},
  \bibinfo {author} {\bibfnamefont {I.}~\bibnamefont {Khan}}, \bibinfo {author}
  {\bibfnamefont {V.}~\bibnamefont {Makarov}}, \bibinfo {author} {\bibfnamefont
  {C.}~\bibnamefont {Marquardt}},\ and\ \bibinfo {author} {\bibfnamefont
  {G.}~\bibnamefont {Leuchs}},\ }\href
  {https://doi.org/10.1109/JSTQE.2014.2365585} {\bibfield  {journal} {\bibinfo
  {journal} {IEEE Journal of Selected Topics in Quantum Electronics}\ }\textbf
  {\bibinfo {volume} {21}},\ \bibinfo {pages} {168} (\bibinfo {year}
  {2014}{\natexlab{a}})}\BibitemShut {NoStop}%
\bibitem [{\citenamefont {Jain}\ \emph
  {et~al.}(2014{\natexlab{b}})\citenamefont {Jain}, \citenamefont {Anisimova},
  \citenamefont {Khan}, \citenamefont {Makarov}, \citenamefont {Marquardt},\
  and\ \citenamefont {Leuchs}}]{jain2014trojan}%
  \BibitemOpen
  \bibfield  {author} {\bibinfo {author} {\bibfnamefont {N.}~\bibnamefont
  {Jain}}, \bibinfo {author} {\bibfnamefont {E.}~\bibnamefont {Anisimova}},
  \bibinfo {author} {\bibfnamefont {I.}~\bibnamefont {Khan}}, \bibinfo {author}
  {\bibfnamefont {V.}~\bibnamefont {Makarov}}, \bibinfo {author} {\bibfnamefont
  {C.}~\bibnamefont {Marquardt}},\ and\ \bibinfo {author} {\bibfnamefont
  {G.}~\bibnamefont {Leuchs}},\ }\href
  {https://doi.org/10.1088/1367-2630/16/12/123030} {\bibfield  {journal}
  {\bibinfo  {journal} {New Journal of Physics}\ }\textbf {\bibinfo {volume}
  {16}},\ \bibinfo {pages} {123030} (\bibinfo {year}
  {2014}{\natexlab{b}})}\BibitemShut {NoStop}%
\bibitem [{\citenamefont {Yoshino}\ \emph {et~al.}(2018)\citenamefont
  {Yoshino}, \citenamefont {Fujiwara}, \citenamefont {Nakata}, \citenamefont
  {Sumiya}, \citenamefont {Sasaki}, \citenamefont {Takeoka}, \citenamefont
  {Sasaki}, \citenamefont {Tajima}, \citenamefont {Koashi},\ and\ \citenamefont
  {Tomita}}]{yoshino2018quantum}%
  \BibitemOpen
  \bibfield  {author} {\bibinfo {author} {\bibfnamefont {K.-i.}\ \bibnamefont
  {Yoshino}}, \bibinfo {author} {\bibfnamefont {M.}~\bibnamefont {Fujiwara}},
  \bibinfo {author} {\bibfnamefont {K.}~\bibnamefont {Nakata}}, \bibinfo
  {author} {\bibfnamefont {T.}~\bibnamefont {Sumiya}}, \bibinfo {author}
  {\bibfnamefont {T.}~\bibnamefont {Sasaki}}, \bibinfo {author} {\bibfnamefont
  {M.}~\bibnamefont {Takeoka}}, \bibinfo {author} {\bibfnamefont
  {M.}~\bibnamefont {Sasaki}}, \bibinfo {author} {\bibfnamefont
  {A.}~\bibnamefont {Tajima}}, \bibinfo {author} {\bibfnamefont
  {M.}~\bibnamefont {Koashi}},\ and\ \bibinfo {author} {\bibfnamefont
  {A.}~\bibnamefont {Tomita}},\ }\href
  {https://doi.org/10.1038/s41534-017-0057-8} {\bibfield  {journal} {\bibinfo
  {journal} {npj Quantum Information}\ }\textbf {\bibinfo {volume} {4}},\
  \bibinfo {pages} {8} (\bibinfo {year} {2018})}\BibitemShut {NoStop}%
\bibitem [{\citenamefont {Roberts}\ \emph {et~al.}(2018)\citenamefont
  {Roberts}, \citenamefont {Pittaluga}, \citenamefont {Minder}, \citenamefont
  {Lucamarini}, \citenamefont {Dynes}, \citenamefont {Yuan},\ and\
  \citenamefont {Shields}}]{roberts2018patterning}%
  \BibitemOpen
  \bibfield  {author} {\bibinfo {author} {\bibfnamefont {G.}~\bibnamefont
  {Roberts}}, \bibinfo {author} {\bibfnamefont {M.}~\bibnamefont {Pittaluga}},
  \bibinfo {author} {\bibfnamefont {M.}~\bibnamefont {Minder}}, \bibinfo
  {author} {\bibfnamefont {M.}~\bibnamefont {Lucamarini}}, \bibinfo {author}
  {\bibfnamefont {J.}~\bibnamefont {Dynes}}, \bibinfo {author} {\bibfnamefont
  {Z.}~\bibnamefont {Yuan}},\ and\ \bibinfo {author} {\bibfnamefont
  {A.}~\bibnamefont {Shields}},\ }\href {https://doi.org/10.1364/OL.43.005110}
  {\bibfield  {journal} {\bibinfo  {journal} {Optics Letters}\ }\textbf
  {\bibinfo {volume} {43}},\ \bibinfo {pages} {5110} (\bibinfo {year}
  {2018})}\BibitemShut {NoStop}%
\bibitem [{\citenamefont {Lu}\ \emph {et~al.}(2021)\citenamefont {Lu},
  \citenamefont {Lin}, \citenamefont {Wang}, \citenamefont {Fan-Yuan},
  \citenamefont {Ye}, \citenamefont {Wang}, \citenamefont {Yin}, \citenamefont
  {He}, \citenamefont {Chen}, \citenamefont {Guo} \emph
  {et~al.}}]{lu2021intensity}%
  \BibitemOpen
  \bibfield  {author} {\bibinfo {author} {\bibfnamefont {F.-Y.}\ \bibnamefont
  {Lu}}, \bibinfo {author} {\bibfnamefont {X.}~\bibnamefont {Lin}}, \bibinfo
  {author} {\bibfnamefont {S.}~\bibnamefont {Wang}}, \bibinfo {author}
  {\bibfnamefont {G.-J.}\ \bibnamefont {Fan-Yuan}}, \bibinfo {author}
  {\bibfnamefont {P.}~\bibnamefont {Ye}}, \bibinfo {author} {\bibfnamefont
  {R.}~\bibnamefont {Wang}}, \bibinfo {author} {\bibfnamefont {Z.-Q.}\
  \bibnamefont {Yin}}, \bibinfo {author} {\bibfnamefont {D.-Y.}\ \bibnamefont
  {He}}, \bibinfo {author} {\bibfnamefont {W.}~\bibnamefont {Chen}}, \bibinfo
  {author} {\bibfnamefont {G.-C.}\ \bibnamefont {Guo}}, \emph {et~al.},\ }\href
  {https://doi.org/10.1038/s41534-021-00418-x} {\bibfield  {journal} {\bibinfo
  {journal} {npj Quantum Information}\ }\textbf {\bibinfo {volume} {7}},\
  \bibinfo {pages} {75} (\bibinfo {year} {2021})}\BibitemShut {NoStop}%
\bibitem [{\citenamefont {Huang}\ \emph {et~al.}(2019)\citenamefont {Huang},
  \citenamefont {Navarrete}, \citenamefont {Sun}, \citenamefont {Chaiwongkhot},
  \citenamefont {Curty},\ and\ \citenamefont {Makarov}}]{huang2019laser}%
  \BibitemOpen
  \bibfield  {author} {\bibinfo {author} {\bibfnamefont {A.}~\bibnamefont
  {Huang}}, \bibinfo {author} {\bibfnamefont {{\'A}.}~\bibnamefont
  {Navarrete}}, \bibinfo {author} {\bibfnamefont {S.-H.}\ \bibnamefont {Sun}},
  \bibinfo {author} {\bibfnamefont {P.}~\bibnamefont {Chaiwongkhot}}, \bibinfo
  {author} {\bibfnamefont {M.}~\bibnamefont {Curty}},\ and\ \bibinfo {author}
  {\bibfnamefont {V.}~\bibnamefont {Makarov}},\ }\href
  {https://doi.org/10.1103/PhysRevApplied.12.064043} {\bibfield  {journal}
  {\bibinfo  {journal} {Physical Review Applied}\ }\textbf {\bibinfo {volume}
  {12}},\ \bibinfo {pages} {064043} (\bibinfo {year} {2019})}\BibitemShut
  {NoStop}%
\bibitem [{\citenamefont {Pang}\ \emph {et~al.}(2020)\citenamefont {Pang},
  \citenamefont {Yang}, \citenamefont {Zhang}, \citenamefont {Dou},
  \citenamefont {Li}, \citenamefont {Gao},\ and\ \citenamefont
  {Jin}}]{pang2020hacking}%
  \BibitemOpen
  \bibfield  {author} {\bibinfo {author} {\bibfnamefont {X.-L.}\ \bibnamefont
  {Pang}}, \bibinfo {author} {\bibfnamefont {A.-L.}\ \bibnamefont {Yang}},
  \bibinfo {author} {\bibfnamefont {C.-N.}\ \bibnamefont {Zhang}}, \bibinfo
  {author} {\bibfnamefont {J.-P.}\ \bibnamefont {Dou}}, \bibinfo {author}
  {\bibfnamefont {H.}~\bibnamefont {Li}}, \bibinfo {author} {\bibfnamefont
  {J.}~\bibnamefont {Gao}},\ and\ \bibinfo {author} {\bibfnamefont {X.-M.}\
  \bibnamefont {Jin}},\ }\href
  {https://doi.org/10.1103/PhysRevApplied.13.034008} {\bibfield  {journal}
  {\bibinfo  {journal} {Physical Review Applied}\ }\textbf {\bibinfo {volume}
  {13}},\ \bibinfo {pages} {034008} (\bibinfo {year} {2020})}\BibitemShut
  {NoStop}%
\bibitem [{\citenamefont {Ye}\ \emph {et~al.}(2023)\citenamefont {Ye},
  \citenamefont {Chen}, \citenamefont {Zhang}, \citenamefont {Lu},
  \citenamefont {Wang}, \citenamefont {Huang}, \citenamefont {Wang},
  \citenamefont {He}, \citenamefont {Yin}, \citenamefont {Guo} \emph
  {et~al.}}]{ye2023induced}%
  \BibitemOpen
  \bibfield  {author} {\bibinfo {author} {\bibfnamefont {P.}~\bibnamefont
  {Ye}}, \bibinfo {author} {\bibfnamefont {W.}~\bibnamefont {Chen}}, \bibinfo
  {author} {\bibfnamefont {G.-W.}\ \bibnamefont {Zhang}}, \bibinfo {author}
  {\bibfnamefont {F.-Y.}\ \bibnamefont {Lu}}, \bibinfo {author} {\bibfnamefont
  {F.-X.}\ \bibnamefont {Wang}}, \bibinfo {author} {\bibfnamefont {G.-Z.}\
  \bibnamefont {Huang}}, \bibinfo {author} {\bibfnamefont {S.}~\bibnamefont
  {Wang}}, \bibinfo {author} {\bibfnamefont {D.-Y.}\ \bibnamefont {He}},
  \bibinfo {author} {\bibfnamefont {Z.-Q.}\ \bibnamefont {Yin}}, \bibinfo
  {author} {\bibfnamefont {G.-C.}\ \bibnamefont {Guo}}, \emph {et~al.},\ }\href
  {https://doi.org/10.1103/PhysRevApplied.19.054052} {\bibfield  {journal}
  {\bibinfo  {journal} {Physical Review Applied}\ }\textbf {\bibinfo {volume}
  {19}},\ \bibinfo {pages} {054052} (\bibinfo {year} {2023})}\BibitemShut
  {NoStop}%
\bibitem [{\citenamefont {Lu}\ \emph {et~al.}(2023{\natexlab{a}})\citenamefont
  {Lu}, \citenamefont {Ye}, \citenamefont {Wang}, \citenamefont {Wang},
  \citenamefont {Yin}, \citenamefont {Wang}, \citenamefont {Huang},
  \citenamefont {Chen}, \citenamefont {He}, \citenamefont {Fan-Yuan} \emph
  {et~al.}}]{lu2023hacking}%
  \BibitemOpen
  \bibfield  {author} {\bibinfo {author} {\bibfnamefont {F.-Y.}\ \bibnamefont
  {Lu}}, \bibinfo {author} {\bibfnamefont {P.}~\bibnamefont {Ye}}, \bibinfo
  {author} {\bibfnamefont {Z.-H.}\ \bibnamefont {Wang}}, \bibinfo {author}
  {\bibfnamefont {S.}~\bibnamefont {Wang}}, \bibinfo {author} {\bibfnamefont
  {Z.-Q.}\ \bibnamefont {Yin}}, \bibinfo {author} {\bibfnamefont
  {R.}~\bibnamefont {Wang}}, \bibinfo {author} {\bibfnamefont {X.-J.}\
  \bibnamefont {Huang}}, \bibinfo {author} {\bibfnamefont {W.}~\bibnamefont
  {Chen}}, \bibinfo {author} {\bibfnamefont {D.-Y.}\ \bibnamefont {He}},
  \bibinfo {author} {\bibfnamefont {G.-J.}\ \bibnamefont {Fan-Yuan}}, \emph
  {et~al.},\ }\href {https://doi.org/10.1364/OPTICA.485389} {\bibfield
  {journal} {\bibinfo  {journal} {Optica}\ }\textbf {\bibinfo {volume} {10}},\
  \bibinfo {pages} {520} (\bibinfo {year} {2023}{\natexlab{a}})}\BibitemShut
  {NoStop}%
\bibitem [{\citenamefont {Curty}\ \emph {et~al.}(2009)\citenamefont {Curty},
  \citenamefont {Moroder}, \citenamefont {Ma},\ and\ \citenamefont
  {L{\"u}tkenhaus}}]{curtyNonPoissonianStatisticsPoissonian2009}%
  \BibitemOpen
  \bibfield  {author} {\bibinfo {author} {\bibfnamefont {M.}~\bibnamefont
  {Curty}}, \bibinfo {author} {\bibfnamefont {T.}~\bibnamefont {Moroder}},
  \bibinfo {author} {\bibfnamefont {X.}~\bibnamefont {Ma}},\ and\ \bibinfo
  {author} {\bibfnamefont {N.}~\bibnamefont {L{\"u}tkenhaus}},\ }\href
  {https://doi.org/10.1364/OL.34.003238} {\bibfield  {journal} {\bibinfo
  {journal} {Optics Letters}\ }\textbf {\bibinfo {volume} {34}},\ \bibinfo
  {pages} {3238} (\bibinfo {year} {2009})}\BibitemShut {NoStop}%
\bibitem [{\citenamefont {Curty}\ \emph
  {et~al.}(2010{\natexlab{a}})\citenamefont {Curty}, \citenamefont {Ma},
  \citenamefont {Qi},\ and\ \citenamefont
  {Moroder}}]{curtyPassiveDecoystateQuantum2010}%
  \BibitemOpen
  \bibfield  {author} {\bibinfo {author} {\bibfnamefont {M.}~\bibnamefont
  {Curty}}, \bibinfo {author} {\bibfnamefont {X.}~\bibnamefont {Ma}}, \bibinfo
  {author} {\bibfnamefont {B.}~\bibnamefont {Qi}},\ and\ \bibinfo {author}
  {\bibfnamefont {T.}~\bibnamefont {Moroder}},\ }\href
  {https://doi.org/10.1103/PhysRevA.81.022310} {\bibfield  {journal} {\bibinfo
  {journal} {Physical Review A}\ }\textbf {\bibinfo {volume} {81}},\ \bibinfo
  {pages} {022310} (\bibinfo {year} {2010}{\natexlab{a}})}\BibitemShut
  {NoStop}%
\bibitem [{\citenamefont {Curty}\ \emph
  {et~al.}(2010{\natexlab{b}})\citenamefont {Curty}, \citenamefont {Ma},
  \citenamefont {Lo},\ and\ \citenamefont
  {L{\"u}tkenhaus}}]{curtyPassiveSourcesBennettBrassard2010}%
  \BibitemOpen
  \bibfield  {author} {\bibinfo {author} {\bibfnamefont {M.}~\bibnamefont
  {Curty}}, \bibinfo {author} {\bibfnamefont {X.}~\bibnamefont {Ma}}, \bibinfo
  {author} {\bibfnamefont {H.-K.}\ \bibnamefont {Lo}},\ and\ \bibinfo {author}
  {\bibfnamefont {N.}~\bibnamefont {L{\"u}tkenhaus}},\ }\href
  {https://doi.org/10.1103/PhysRevA.82.052325} {\bibfield  {journal} {\bibinfo
  {journal} {Physical Review A}\ }\textbf {\bibinfo {volume} {82}},\ \bibinfo
  {pages} {052325} (\bibinfo {year} {2010}{\natexlab{b}})}\BibitemShut
  {NoStop}%
\bibitem [{\citenamefont {Hwang}(2003)}]{hwang2003quantum}%
  \BibitemOpen
  \bibfield  {author} {\bibinfo {author} {\bibfnamefont {W.-Y.}\ \bibnamefont
  {Hwang}},\ }\href {https://doi.org/10.1103/PhysRevLett.91.057901} {\bibfield
  {journal} {\bibinfo  {journal} {Physical Review Letters}\ }\textbf {\bibinfo
  {volume} {91}},\ \bibinfo {pages} {057901} (\bibinfo {year}
  {2003})}\BibitemShut {NoStop}%
\bibitem [{\citenamefont {Lo}\ \emph {et~al.}(2005)\citenamefont {Lo},
  \citenamefont {Ma},\ and\ \citenamefont {Chen}}]{lo2005decoy}%
  \BibitemOpen
  \bibfield  {author} {\bibinfo {author} {\bibfnamefont {H.-K.}\ \bibnamefont
  {Lo}}, \bibinfo {author} {\bibfnamefont {X.}~\bibnamefont {Ma}},\ and\
  \bibinfo {author} {\bibfnamefont {K.}~\bibnamefont {Chen}},\ }\href
  {https://doi.org/10.1103/PhysRevLett.94.230504} {\bibfield  {journal}
  {\bibinfo  {journal} {Physical Review Letters}\ }\textbf {\bibinfo {volume}
  {94}},\ \bibinfo {pages} {230504} (\bibinfo {year} {2005})}\BibitemShut
  {NoStop}%
\bibitem [{\citenamefont {Wang}(2005)}]{wang2005beating}%
  \BibitemOpen
  \bibfield  {author} {\bibinfo {author} {\bibfnamefont {X.-B.}\ \bibnamefont
  {Wang}},\ }\href {https://doi.org/10.1103/PhysRevLett.94.230503} {\bibfield
  {journal} {\bibinfo  {journal} {Physical Review Letters}\ }\textbf {\bibinfo
  {volume} {94}},\ \bibinfo {pages} {230503} (\bibinfo {year}
  {2005})}\BibitemShut {NoStop}%
\bibitem [{\citenamefont {Curty}\ \emph {et~al.}(2015)\citenamefont {Curty},
  \citenamefont {Jofre}, \citenamefont {Pruneri},\ and\ \citenamefont
  {Mitchell}}]{curty2015passive}%
  \BibitemOpen
  \bibfield  {author} {\bibinfo {author} {\bibfnamefont {M.}~\bibnamefont
  {Curty}}, \bibinfo {author} {\bibfnamefont {M.}~\bibnamefont {Jofre}},
  \bibinfo {author} {\bibfnamefont {V.}~\bibnamefont {Pruneri}},\ and\ \bibinfo
  {author} {\bibfnamefont {M.~W.}\ \bibnamefont {Mitchell}},\ }\href
  {https://doi.org/10.3390/e17064064} {\bibfield  {journal} {\bibinfo
  {journal} {Entropy}\ }\textbf {\bibinfo {volume} {17}},\ \bibinfo {pages}
  {4064} (\bibinfo {year} {2015})}\BibitemShut {NoStop}%
\bibitem [{\citenamefont {Wang}\ \emph {et~al.}(2016)\citenamefont {Wang},
  \citenamefont {Zhang},\ and\ \citenamefont {Wang}}]{wang2016scheme}%
  \BibitemOpen
  \bibfield  {author} {\bibinfo {author} {\bibfnamefont {Q.}~\bibnamefont
  {Wang}}, \bibinfo {author} {\bibfnamefont {C.-H.}\ \bibnamefont {Zhang}},\
  and\ \bibinfo {author} {\bibfnamefont {X.-B.}\ \bibnamefont {Wang}},\ }\href
  {https://doi.org/10.1103/PhysRevA.93.032312} {\bibfield  {journal} {\bibinfo
  {journal} {Physical Review A}\ }\textbf {\bibinfo {volume} {93}},\ \bibinfo
  {pages} {032312} (\bibinfo {year} {2016})}\BibitemShut {NoStop}%
\bibitem [{\citenamefont {Zhang}\ \emph {et~al.}(2018)\citenamefont {Zhang},
  \citenamefont {Wang}, \citenamefont {Zhou}, \citenamefont {Wang},
  \citenamefont {Zhang}, \citenamefont {Yin}, \citenamefont {Chen},
  \citenamefont {Han}, \citenamefont {Guo},\ and\ \citenamefont
  {Wang}}]{zhang2018proof}%
  \BibitemOpen
  \bibfield  {author} {\bibinfo {author} {\bibfnamefont {C.-H.}\ \bibnamefont
  {Zhang}}, \bibinfo {author} {\bibfnamefont {D.}~\bibnamefont {Wang}},
  \bibinfo {author} {\bibfnamefont {X.-Y.}\ \bibnamefont {Zhou}}, \bibinfo
  {author} {\bibfnamefont {S.}~\bibnamefont {Wang}}, \bibinfo {author}
  {\bibfnamefont {L.-B.}\ \bibnamefont {Zhang}}, \bibinfo {author}
  {\bibfnamefont {Z.-Q.}\ \bibnamefont {Yin}}, \bibinfo {author} {\bibfnamefont
  {W.}~\bibnamefont {Chen}}, \bibinfo {author} {\bibfnamefont {Z.-F.}\
  \bibnamefont {Han}}, \bibinfo {author} {\bibfnamefont {G.-C.}\ \bibnamefont
  {Guo}},\ and\ \bibinfo {author} {\bibfnamefont {Q.}~\bibnamefont {Wang}},\
  }\href {https://doi.org/10.1364/OE.26.025921} {\bibfield  {journal} {\bibinfo
   {journal} {Optics Express}\ }\textbf {\bibinfo {volume} {26}},\ \bibinfo
  {pages} {25921} (\bibinfo {year} {2018})}\BibitemShut {NoStop}%
\bibitem [{\citenamefont {Wang}\ \emph {et~al.}(2023)\citenamefont {Wang},
  \citenamefont {Wang}, \citenamefont {Hu}, \citenamefont {Zapatero},
  \citenamefont {Qian}, \citenamefont {Qi}, \citenamefont {Curty},\ and\
  \citenamefont {Lo}}]{wangFullyPassiveQuantum2023}%
  \BibitemOpen
  \bibfield  {author} {\bibinfo {author} {\bibfnamefont {W.}~\bibnamefont
  {Wang}}, \bibinfo {author} {\bibfnamefont {R.}~\bibnamefont {Wang}}, \bibinfo
  {author} {\bibfnamefont {C.}~\bibnamefont {Hu}}, \bibinfo {author}
  {\bibfnamefont {V.}~\bibnamefont {Zapatero}}, \bibinfo {author}
  {\bibfnamefont {L.}~\bibnamefont {Qian}}, \bibinfo {author} {\bibfnamefont
  {B.}~\bibnamefont {Qi}}, \bibinfo {author} {\bibfnamefont {M.}~\bibnamefont
  {Curty}},\ and\ \bibinfo {author} {\bibfnamefont {H.-K.}\ \bibnamefont
  {Lo}},\ }\href {https://doi.org/10.1103/PhysRevLett.130.220801} {\bibfield
  {journal} {\bibinfo  {journal} {Physical Review Letters}\ }\textbf {\bibinfo
  {volume} {130}},\ \bibinfo {pages} {220801} (\bibinfo {year}
  {2023})}\BibitemShut {NoStop}%
\bibitem [{\citenamefont {Zapatero}\ \emph {et~al.}(2023)\citenamefont
  {Zapatero}, \citenamefont {Wang},\ and\ \citenamefont
  {Curty}}]{zapateroFullyPassiveTransmitter2023}%
  \BibitemOpen
  \bibfield  {author} {\bibinfo {author} {\bibfnamefont {V.}~\bibnamefont
  {Zapatero}}, \bibinfo {author} {\bibfnamefont {W.}~\bibnamefont {Wang}},\
  and\ \bibinfo {author} {\bibfnamefont {M.}~\bibnamefont {Curty}},\ }\href
  {https://doi.org/10.1088/2058-9565/acbc46} {\bibfield  {journal} {\bibinfo
  {journal} {Quantum Science and Technology}\ }\textbf {\bibinfo {volume}
  {8}},\ \bibinfo {pages} {025014} (\bibinfo {year} {2023})}\BibitemShut
  {NoStop}%
\bibitem [{\citenamefont {Kang}\ \emph {et~al.}(2023)\citenamefont {Kang},
  \citenamefont {Lu}, \citenamefont {Wang}, \citenamefont {Chen}, \citenamefont
  {Wang}, \citenamefont {Yin}, \citenamefont {He}, \citenamefont {Chen},
  \citenamefont {Fan-Yuan}, \citenamefont {Guo} \emph
  {et~al.}}]{kang2023patterning}%
  \BibitemOpen
  \bibfield  {author} {\bibinfo {author} {\bibfnamefont {X.}~\bibnamefont
  {Kang}}, \bibinfo {author} {\bibfnamefont {F.-Y.}\ \bibnamefont {Lu}},
  \bibinfo {author} {\bibfnamefont {S.}~\bibnamefont {Wang}}, \bibinfo {author}
  {\bibfnamefont {J.-L.}\ \bibnamefont {Chen}}, \bibinfo {author}
  {\bibfnamefont {Z.-H.}\ \bibnamefont {Wang}}, \bibinfo {author}
  {\bibfnamefont {Z.-Q.}\ \bibnamefont {Yin}}, \bibinfo {author} {\bibfnamefont
  {D.-Y.}\ \bibnamefont {He}}, \bibinfo {author} {\bibfnamefont
  {W.}~\bibnamefont {Chen}}, \bibinfo {author} {\bibfnamefont {G.-J.}\
  \bibnamefont {Fan-Yuan}}, \bibinfo {author} {\bibfnamefont {G.-C.}\
  \bibnamefont {Guo}}, \emph {et~al.},\ }\href
  {https://opg.optica.org/jlt/abstract.cfm?URI=jlt-41-1-75} {\bibfield
  {journal} {\bibinfo  {journal} {Journal of Lightwave Technology}\ }\textbf
  {\bibinfo {volume} {41}},\ \bibinfo {pages} {75} (\bibinfo {year}
  {2023})}\BibitemShut {NoStop}%
\bibitem [{\citenamefont {Lu}\ \emph {et~al.}(2023{\natexlab{b}})\citenamefont
  {Lu}, \citenamefont {Wang}, \citenamefont {Wang}, \citenamefont {Yin},
  \citenamefont {Chen}, \citenamefont {Kang}, \citenamefont {He}, \citenamefont
  {Chen}, \citenamefont {Fan-Yuan}, \citenamefont {Guo},\ and\ \citenamefont
  {Han}}]{lu2023intensity}%
  \BibitemOpen
  \bibfield  {author} {\bibinfo {author} {\bibfnamefont {F.-Y.}\ \bibnamefont
  {Lu}}, \bibinfo {author} {\bibfnamefont {Z.-H.}\ \bibnamefont {Wang}},
  \bibinfo {author} {\bibfnamefont {S.}~\bibnamefont {Wang}}, \bibinfo {author}
  {\bibfnamefont {Z.-Q.}\ \bibnamefont {Yin}}, \bibinfo {author} {\bibfnamefont
  {J.-L.}\ \bibnamefont {Chen}}, \bibinfo {author} {\bibfnamefont
  {X.}~\bibnamefont {Kang}}, \bibinfo {author} {\bibfnamefont {D.-Y.}\
  \bibnamefont {He}}, \bibinfo {author} {\bibfnamefont {W.}~\bibnamefont
  {Chen}}, \bibinfo {author} {\bibfnamefont {G.-J.}\ \bibnamefont {Fan-Yuan}},
  \bibinfo {author} {\bibfnamefont {G.-C.}\ \bibnamefont {Guo}},\ and\ \bibinfo
  {author} {\bibfnamefont {Z.-F.}\ \bibnamefont {Han}},\ }\href
  {https://doi.org/10.1109/JLT.2023.3247766} {\bibfield  {journal} {\bibinfo
  {journal} {Journal of Lightwave Technology}\ }\textbf {\bibinfo {volume}
  {41}},\ \bibinfo {pages} {4895} (\bibinfo {year}
  {2023}{\natexlab{b}})}\BibitemShut {NoStop}%
\bibitem [{\citenamefont {Zhou}\ \emph {et~al.}(2016)\citenamefont {Zhou},
  \citenamefont {Yu},\ and\ \citenamefont {Wang}}]{zhou2016making}%
  \BibitemOpen
  \bibfield  {author} {\bibinfo {author} {\bibfnamefont {Y.-H.}\ \bibnamefont
  {Zhou}}, \bibinfo {author} {\bibfnamefont {Z.-W.}\ \bibnamefont {Yu}},\ and\
  \bibinfo {author} {\bibfnamefont {X.-B.}\ \bibnamefont {Wang}},\ }\href
  {https://doi.org/10.1103/PhysRevA.93.042324} {\bibfield  {journal} {\bibinfo
  {journal} {Physical Review A}\ }\textbf {\bibinfo {volume} {93}},\ \bibinfo
  {pages} {042324} (\bibinfo {year} {2016})}\BibitemShut {NoStop}%
\bibitem [{\citenamefont {Yu}\ \emph {et~al.}(2013)\citenamefont {Yu},
  \citenamefont {Zhou},\ and\ \citenamefont {Wang}}]{yu2013three}%
  \BibitemOpen
  \bibfield  {author} {\bibinfo {author} {\bibfnamefont {Z.-W.}\ \bibnamefont
  {Yu}}, \bibinfo {author} {\bibfnamefont {Y.-H.}\ \bibnamefont {Zhou}},\ and\
  \bibinfo {author} {\bibfnamefont {X.-B.}\ \bibnamefont {Wang}},\ }\href
  {https://doi.org/10.1103/PhysRevA.88.062339} {\bibfield  {journal} {\bibinfo
  {journal} {Physical Review A}\ }\textbf {\bibinfo {volume} {88}},\ \bibinfo
  {pages} {062339} (\bibinfo {year} {2013})}\BibitemShut {NoStop}%
\bibitem [{\citenamefont {Ma}\ and\ \citenamefont
  {Razavi}(2012)}]{ma2012alternative}%
  \BibitemOpen
  \bibfield  {author} {\bibinfo {author} {\bibfnamefont {X.}~\bibnamefont
  {Ma}}\ and\ \bibinfo {author} {\bibfnamefont {M.}~\bibnamefont {Razavi}},\
  }\href {https://doi.org/10.1103/PhysRevA.86.062319} {\bibfield  {journal}
  {\bibinfo  {journal} {Physical Review A}\ }\textbf {\bibinfo {volume} {86}},\
  \bibinfo {pages} {062319} (\bibinfo {year} {2012})}\BibitemShut {NoStop}%
\bibitem [{\citenamefont {Xu}\ \emph {et~al.}(2013)\citenamefont {Xu},
  \citenamefont {Curty}, \citenamefont {Qi},\ and\ \citenamefont
  {Lo}}]{xu2013practical}%
  \BibitemOpen
  \bibfield  {author} {\bibinfo {author} {\bibfnamefont {F.}~\bibnamefont
  {Xu}}, \bibinfo {author} {\bibfnamefont {M.}~\bibnamefont {Curty}}, \bibinfo
  {author} {\bibfnamefont {B.}~\bibnamefont {Qi}},\ and\ \bibinfo {author}
  {\bibfnamefont {H.-K.}\ \bibnamefont {Lo}},\ }\href
  {https://doi.org/10.1088/1367-2630/15/11/113007} {\bibfield  {journal}
  {\bibinfo  {journal} {New Journal of Physics}\ }\textbf {\bibinfo {volume}
  {15}},\ \bibinfo {pages} {113007} (\bibinfo {year} {2013})}\BibitemShut
  {NoStop}%
\bibitem [{\citenamefont {Xu}\ \emph {et~al.}(2014)\citenamefont {Xu},
  \citenamefont {Xu},\ and\ \citenamefont {Lo}}]{xu2014protocol}%
  \BibitemOpen
  \bibfield  {author} {\bibinfo {author} {\bibfnamefont {F.}~\bibnamefont
  {Xu}}, \bibinfo {author} {\bibfnamefont {H.}~\bibnamefont {Xu}},\ and\
  \bibinfo {author} {\bibfnamefont {H.-K.}\ \bibnamefont {Lo}},\ }\href
  {https://doi.org/10.1103/PhysRevA.89.052333} {\bibfield  {journal} {\bibinfo
  {journal} {Physical Review A}\ }\textbf {\bibinfo {volume} {89}},\ \bibinfo
  {pages} {052333} (\bibinfo {year} {2014})}\BibitemShut {NoStop}%
\bibitem [{\citenamefont {Jiang}\ \emph {et~al.}(2021)\citenamefont {Jiang},
  \citenamefont {Yu}, \citenamefont {Hu},\ and\ \citenamefont
  {Wang}}]{jiang2021higher}%
  \BibitemOpen
  \bibfield  {author} {\bibinfo {author} {\bibfnamefont {C.}~\bibnamefont
  {Jiang}}, \bibinfo {author} {\bibfnamefont {Z.-W.}\ \bibnamefont {Yu}},
  \bibinfo {author} {\bibfnamefont {X.-L.}\ \bibnamefont {Hu}},\ and\ \bibinfo
  {author} {\bibfnamefont {X.-B.}\ \bibnamefont {Wang}},\ }\href
  {https://doi.org/10.1103/PhysRevA.103.012402} {\bibfield  {journal} {\bibinfo
   {journal} {Physical Review A}\ }\textbf {\bibinfo {volume} {103}},\ \bibinfo
  {pages} {012402} (\bibinfo {year} {2021})}\BibitemShut {NoStop}%
\end{thebibliography}%

\end{document}